\documentclass[11pt]{article}
\usepackage{jheppub}

\usepackage{amsmath}
\usepackage{amssymb}
\usepackage{verbatim}

\usepackage[T1,OT1]{fontenc}

\usepackage{graphicx}

\allowdisplaybreaks[4]

\usepackage[hang]{footmisc}
\renewcommand{\d}{\mathrm{d}}
\newcommand{\e}{\mathrm{e}}

\newcommand{\nl}{\notag \\ &\quad\,}
\newcommand{\nll}{\notag \\ &}

\title{Topological Constraints in the LARGE-Volume Scenario}

\author{Daniel Junghans}

\affiliation{Harvard University, Center of Mathematical Sciences and Applications, 20 Garden Street,\\Cambridge, MA 02138, USA}

\emailAdd{djunghans @ cmsa.fas.harvard.edu}

\notoc

\abstract{
We elaborate on recent results regarding the self-consistency of de Sitter vacua in the LARGE-volume scenario of type IIB string theory. In particular, we analyze to what extent the control over warping, curvature and $g_s$ corrections depends on the topology and the orientifold/brane data of a compactification.
We compute a general bound on the magnitude of these corrections which strongly constrains the D3 tadpole.
The minimally required tadpole ranges from $\mathcal{O}(500)$ to $\mathcal{O}(10^6)$ or more and depends strongly on other data, in particular on the Euler number of the Calabi-Yau 3-fold, the triple-self-intersection and Euler numbers of the small divisor and the coefficient $a_s$ appearing in the non-perturbative superpotential.
We give arguments suggesting that satisfying these constraints is very challenging and perhaps impossible.
}

\begin{document}

\numberwithin{equation}{section}

\maketitle

\newpage

\section{Introduction}

Whether string theory has dS vacua has been debated for almost two decades since the seminal work \cite{Kachru:2003aw}. However, in spite of a huge amount of literature on the subject,
so far no consensus about the status of dS has been reached in the community (see \cite{Danielsson:2018ztv} for a review).
A new spin was recently put on this question by swampland conjectures such as \cite{Obied:2018sgi} and variants thereof, which posit that dS space is ruled out or at least strongly constrained in quantum gravity.
If this is true, it would have far-reaching consequences for phenomenology, in particular for the nature of dark energy. It is therefore important to settle this debate,
either by constructing explicit dS examples that are beyond suspicion or by showing in general that the proposed constructions must fail.

In this paper, we will study the self-consistency of the LARGE-Volume Scenario (LVS) \cite{Balasubramanian:2005zx, Conlon:2005ki}, which is one of the leading proposals for dS vacua in string theory.
A common lore is that the LVS is extremely well controlled since the exponentially large volume should strongly suppress any unwanted corrections. However, it was recently shown in \cite{Junghans:2022exo} that this is not the case if one considers dS vacua instead of AdS vacua. Indeed, LVS dS vacua are vulnerable to a number of potentially large warping, curvature and $g_s$ corrections. These corrections can change the sign of the vacuum energy and the eigenvalues of the mass matrix and thus put candidate dS vacua in the swampland.

Specifically, \cite{Junghans:2022exo} characterized a number of relevant corrections
in terms of control parameters $\lambda_i$,
which were defined in such a way that self-consistent vacua have to satisfy $\lambda_i\ll 1$.
These conditions were then studied in an explicit Calabi-Yau (CY) compactification constructed in \cite{Crino:2020qwk}, which in principle has the right ingredients to admit LVS dS vacua. However, the analysis of \cite{Junghans:2022exo} revealed that it is impossible to satisfy all conditions at the same time whenever a positive vacuum energy is imposed.
Instead, at least one of the $\lambda_i$ parameters is always $\mathcal{O}(1)$ in every candidate dS vacuum of the model, implying large corrections and a loss of control.

A natural question is whether better controlled dS vacua can be obtained in other LVS models. In particular, one may wonder to what extent the $\mathcal{O}(1)$ bound on the control parameters can be relaxed
by changing data such as the topological invariants of the CY, the orientifold involution or the non-perturbative effect. In this note, we provide an answer to this question by generalizing the bound of \cite{Junghans:2022exo} to compactifications with arbitrary topologies and orientifold/brane data (assuming the minimal scenario with $h_+^{1,1}=2$ and an uplift generated by an anti-D3 brane).
In particular, we derive a bound on the control parameters which depends on the D3 tadpole, the Euler number of the CY, the triple-self-intersection and Euler numbers of the small divisor and the coefficient $a_s$ appearing in the non-perturbative superpotential.
This allows us to precisely determine which values these numbers must take in order to admit dS vacua with small corrections. In particular, our bound puts strong constraints on the D3 tadpole. We find that the minimally required tadpole can range from $\mathcal{O}(500)$ to $\mathcal{O}(10^6)$ or more, where the precise value depends strongly on the other topological data and on the desired degree of control over the corrections.

Our work is also inspired by the recent work \cite{Gao:2022fdi}, which focussed on a particular combination of the control parameters studied in \cite{Junghans:2022exo} and showed that it becomes small in models with sufficiently large D3 tadpoles. This constraint was called the ``LVS parametric tadpole constraint'' (PTC). Motivated by their observation, \cite{Gao:2022fdi} proposed that the main obstacle in the quest for controlled LVS dS vacua lies in finding models with a large enough tadpole to satisfy the PTC.
However, as we show in this work,
this is too optimistic. Although we find---in agreement with the idea of \cite{Gao:2022fdi}---that large tadpoles indeed help to suppress the corrections, the PTC taken alone is a rather weak constraint which is in fact much easier to satisfy than the full set of constraints arising from the corrections studied in
\cite{Junghans:2022exo}. In particular, the PTC suggests excellent control already in models with an $\mathcal{O}(100)$ tadpole, whereas tadpoles that are several orders of magnitude larger can be required if the more general constraints of \cite{Junghans:2022exo} are imposed.

This paper is organized as follows. In Section \ref{sec:review}, we briefly review some useful formulae of the LVS and the key results of \cite{Junghans:2022exo, Gao:2022fdi}. In Section \ref{sec:bound}, we derive a bound on the magnitude of corrections in LVS dS vacua. In Section \ref{sec:disc}, we discuss the implications of the bound and provide a number of illustrative plots. We also argue that constructing LVS models satisfying our constraints may not be possible and discuss several effects that further increase the required tadpoles beyond our bound. We conclude in Section \ref{sec:concl} with a summary of our results. In Appendix \ref{sec:ptc}, we compare our conventions for the anti-brane uplift and the control parameters to those of \cite{Gao:2022fdi}. We also give a general argument that the uplift term depends on a model-dependent constant which is not determined by the local solution in the throat, and we show that this ambiguity does not affect our bound.

\section{Review of previous results}
\label{sec:review}

\subsection{LVS dS vacua}
\label{sec:review1}

We will focus on the minimal LVS setup with $h^{1,1}=h^{1,1}_+=2$ and a dS uplift generated by an anti-D3 brane in a warped throat.
We refer to \cite{Balasubramanian:2005zx, Conlon:2005ki,Crino:2020qwk, Junghans:2022exo} for details about the construction and only collect some useful formulae here. Our starting point is
the $F$-term potential for the two K\"ahler moduli $\tau_b$, $\tau_s$ and the conifold modulus $\zeta$:\footnote{See \cite{Douglas:2007tu, Douglas:2008jx, Bena:2018fqc, Blumenhagen:2019qcg, Dudas:2019pls} for the derivation of the conifold part of the potential and Section 2 of \cite{Junghans:2022exo} for a discussion of caveats related to this term.
}
\begin{align}
V &= \frac{4a_s^2|A_s|^2g_s\sqrt{\tau_s}\e^{-2a_s(\tau_s-\frac{\chi_s}{24g_s})}}{3\kappa_s \mathcal{V}} - \frac{2a_s|A_s|g_s\tau_s|W_0|\e^{-a_s(\tau_s-\frac{\chi_s}{24g_s})}}{\mathcal{V}^2} + \frac{3|W_0|^2\hat\xi}{8\sqrt{g_s}\mathcal{V}^3} \nl + \frac{\zeta^{4/3}}{2\pi^2 c^\prime \mathcal{V}^{4/3}} \left( \frac{\pi c^\prime c^{\prime\prime}}{g_sM^2} + \frac{\pi^2 K^2}{g_s^2M^2} + \frac{\pi K}{g_sM}\ln\zeta + \frac{1}{4}\ln^2\zeta \right), \label{lvs-potential}
\end{align}
where $c^\prime\approx 1.18$, $c^{\prime\prime}\approx 1.75$ and we will set $A_s=1$ in the following.
Furthermore, $\mathcal{V}=\tau_b^{3/2}-\kappa_s\tau_s^{3/2}$ is the CY volume, where $\kappa_s= \frac{\sqrt{2}}{3\sqrt{\kappa_{sss}}}$ and $\kappa_{sss}$ is the triple-self-intersection number of the small divisor. The various other constants are defined as follows: $g_s$ is the string coupling, $W_0$ is the flux superpotential, $K$ and $M$ are flux numbers characterizing the conifold region and $\chi_s$ is the Euler number of the small divisor.
Furthermore, $a_s=\frac{2\pi}{N_C}$, where $N_C$ is either the dual Coxeter number of the gauge group on the D7-brane stack providing the non-perturbative effect or $N_C=1$ if the non-perturbative effect is generated by D3 instantons. Finally, $\hat\xi$ is related to an orientifold-corrected Euler number $\hat\chi$ of the CY 3-fold \cite{Becker:2002nn, Minasian:2015bxa} with
\begin{equation}
\hat\xi= -\frac{\hat\chi\zeta(3)}{2(2\pi)^3}, \qquad \hat\chi = \chi + 2 \int_X D_\text{O7}^3,
\end{equation}
where $\zeta(3)\approx 1.20$, $\chi$ is the ordinary Euler number and $D_\text{O7}$ is the Poincar\'e dual of the divisor wrapped by the O7 planes.

The equations of motion are solved for
\begin{align}
\mathcal{V} &= \frac{a_s\tau_s-1}{4a_s\tau_s-1}\frac{3\kappa_s|W_0| \sqrt{\tau_s}}{a_s|A_s|}\, \e^{a_s(\tau_s-\frac{\chi_s}{24g_s})}, \label{tb0} \\
\tau_s &= \frac{\hat\xi^{2/3}}{(2\kappa_s)^{2/3}g_s} + \frac{1}{3a_s} + \frac{4\alpha}{15a_s} + \mathcal{O}(g_s), \label{ts0} \\
\zeta &= \e^{-\frac{2\pi K}{g_sM}-\frac{3}{4}+\sqrt{\frac{9}{16}-\frac{4\pi c^{\prime}c^{\prime\prime}}{g_sM^2}}}, \label{zeta0}
\end{align}
where $g_sM^2$ has to satisfy
\begin{equation}
g_sM^2\ge \frac{64\pi c^{\prime}c^{\prime\prime}}{9}\approx 46.1 \label{gsm}
\end{equation}
and the uplift parameter $\alpha$ is defined as
\begin{equation}
\alpha = \frac{20a_s q_0 \zeta^{4/3} \mathcal{V}^{5/3}}{27g_s|W_0|^2\kappa_s \sqrt{\tau_s}} \label{alpha}
\end{equation}
with
\begin{equation}
q_0 = \frac{3}{32\pi^2 c^\prime} \left(3-\sqrt{9-\frac{64\pi c^\prime c^{\prime\prime}}{g_sM^2}}\right). \label{q0}
\end{equation}

The vacuum energy is
\begin{equation}
V_0 = \frac{3g_s\kappa_s|W_0|^2\sqrt{\tau_s}}{8\mathcal{V}^3a_s}\left( \alpha-1+ \mathcal{O}(g_s)\right), \label{v0}
\end{equation}
and the eigenvalues of the Hessian are
\begin{align}
m_1^2 &= \frac{3}{32\pi^2 c^\prime} \sqrt{9-\frac{64\pi c^\prime c^{\prime\prime}}{g_sM^2}} \frac{4\zeta^{2/3}}{9c^\prime g_sM^2 \mathcal{V}^{2/3}},
\qquad m_2^2 = \frac{2g_s|W_0|^2a_s^2\tau_s^2}{\mathcal{V}^{2}}, \notag \\ m_3^2 &= 
\frac{9g_s|W_0|^2\kappa_s\sqrt{\tau_s}}{4a_s \mathcal{V}^{3}} \left(\frac{9}{4}- \alpha + \mathcal{O}(g_s)\right) \label{abc}
\end{align}
up to small corrections. We thus see that, for small $g_s$, dS vacua are obtained in the interval
\begin{equation}
\alpha \in \Big]1,\frac{9}{4}\Big[. \label{range}
\end{equation}

An explicit compactification realizing the above scenario was constructed in \cite{Crino:2020qwk} (based on a CY manifold in the database of \cite{Altman:2014bfa}). The model satisfies
\begin{equation}
Q_3=149, \qquad \hat\xi = \frac{224\zeta(3)}{2(2\pi)^3}, \qquad \kappa_s= \frac{\sqrt{2}}{9}, \qquad \chi_s=3, \qquad a_s=\frac{\pi}{3}, \label{crino}
\end{equation}
where $Q_3$ is the D3 tadpole.

\subsection{Corrections}

It was shown in \cite{Junghans:2022exo} that the potential \eqref{lvs-potential} receives a variety of corrections that potentially invalidate the dS claim. In particular, the following problems were observed:
\begin{itemize}
\item One-loop redefinitions of the small K\"ahler modulus \cite{Conlon:2009xf, Conlon:2009kt, Conlon:2010ji} and D7-brane/O7-plane backreaction corrections \cite{Minasian:2015bxa, Antoniadis:2019rkh} appear with a $1/g_s$ scaling in the $\tau_s$ vev and thus blow up at small coupling.
The volume $\mathcal{V}$ and the uplift parameter $\alpha$ (which determines whether a solution is a dS vacuum according to \eqref{range})
are exponentially sensitive to these corrections \cite{Junghans:2022exo} unless
\begin{equation}
\frac{\hat\xi^{2/3} a_s^2 |\mathcal{C}_s^\text{log}|}{(2\kappa_s)^{2/3}},\frac{2 \hat\xi^{1/3} a_s^2 |\mathcal{C}_1^\xi|}{3(2\kappa_s)^{4/3}}\ll g_s, \qquad
\frac{2a_s|\mathcal{C}_2^\xi|}{3(2\kappa_s)^{2/3}\hat\xi^{1/3}} \ll 1. \label{logxi}
\end{equation}
Here $\mathcal{C}_s^\text{log}$, $\mathcal{C}_i^\xi$ are model-dependent numerical coefficients.

The constraints \eqref{logxi} arise as follows.
The $\mathcal{C}_s^\text{log}$ correction is due to a field redefinition of the form $\tau_s \to \tau_s + \mathcal{C}_s^\text{log} \ln\mathcal{V}$, which was shown to occur for blow-up moduli in several orbifold models in \cite{Conlon:2009xf, Conlon:2009kt} (see also \cite{Conlon:2010ji, Grimm:2017pid, Weissenbacher:2019mef, Weissenbacher:2020cyf, Klaewer:2020lfg} for further arguments). This yields a correction to the K\"ahler potential after expressing it in terms of the redefined K\"ahler coordinate. As a consequence, the moduli vevs and in particular the volume receive corrections as well, which implies that the first inequality in \eqref{logxi} has to hold.

The $\mathcal{C}_i^\xi$ corrections take a similar form in the K\"ahler potential but have a different physical origin. They arise from a curvature correction $\propto \int_X \e^{2A-3\phi/2} c_3(X)$ to the Einstein-Hilbert term, where $c_3(X)$ is the third Chern class and $\e^{2A}$ is the (Einstein-frame) warp factor. For constant warp factor and dilaton, the integral is proportional to the Euler number of the CY 3-fold (or, more precisely, an appropriate generalization thereof for orientifolds \cite{Minasian:2015bxa}) and yields the well-known BBHL term \cite{Becker:2002nn}. However, in many models, the dilaton is not constant due to the backreaction of the D7 branes and O7 planes so that the integral is no longer proportional to the Euler number. This is in particular true when the D7 branes and O7 planes do not lie on top of each other. The BBHL term then receives corrections at the relative order $g_s$, as (very briefly) discussed in F-theory language around (3.9) in \cite{Minasian:2015bxa} (see also Section 3.1.3 and the discussion around (3.15) in \cite{Junghans:2022exo} for the analogous type IIB argument). The corrections were also established in a direct string-theory computation in \cite{Antoniadis:2019rkh}, where it was shown that they can depend logarithmically on the volume in the K\"ahler potential.
As shown in \cite{Junghans:2022exo}, the corrections to the K\"ahler potential yield corrections to the moduli vevs, which implies that the second and third inequalities in \eqref{logxi} have to hold.

Since the numbers on the left-hand sides of the inequalities in \eqref{logxi} are expected to be $\mathcal{O}(1)$ generically, this is a potential problem for the self-consistency of dS vacua (and, possibly, even AdS vacua) in the LVS. One therefore has to identify models where these corrections are absent or where at least the numerical coefficients are much smaller than 1. It is not clear to us whether there could be mechanisms to engineer such a behavior, perhaps similar to known mechanisms engineering a small $W_0$ \cite{Demirtas:2019sip}. We will not study this problem further in this work but stress that addressing it presents a non-trivial obstacle in any attempt to construct explicit LVS dS vacua.

\item A number of further warping, curvature and $g_s$ corrections can be made small individually by appropriate parameter choices (e.g., at sufficiently large volume and small $g_s$), but it turns out that making \emph{all} of them small while at the same time imposing the dS condition \eqref{range} is very difficult.
In order to self-consistently neglect these corrections in the vacuum energy and the moduli masses, one requires\footnote{The corrections appear with numerical coefficients whose precise values are in general unknown and assumed to be $\mathcal{O}(1)$ here.} \cite{Junghans:2022exo}
\begin{equation}
\lambda_i \ll 1
\label{lc}
\end{equation}
with
\begin{align}
\lambda_1&\equiv \text{max}\left(\frac{10+\alpha+4\alpha^2}{30|\alpha-1|}, \frac{135-16\alpha+16\alpha^2}{120|\frac{9}{4}-\alpha|} \right) \frac{(2\kappa_s)^{2/3}}{a_s\hat\xi^{2/3}} g_s, \label{lambda00} \\
\lambda_2 &\equiv \text{max}\left[\frac{1}{|\alpha-1|} \left|7-\frac{15}{\sqrt{9-\frac{64\pi c^\prime c^{\prime\prime}}{g_sM^2}}}\right|,\frac{1}{|\frac{9}{4}-\alpha|} \left|10-\frac{15}{\sqrt{9-\frac{64\pi c^\prime c^{\prime\prime}}{g_sM^2}}}\right| \right] \nl \times \frac{8c^\prime a_s}{9(2\kappa_s)^{2/3}\hat\xi^{1/3}} \frac{g_s^{5/2}M^2\zeta^{2/3}}{\mathcal{V}^{1/3}}, \\
\lambda_3 &\equiv \text{max}\left(\frac{20}{|\alpha-1|},\frac{55}{|\frac{9}{4}-\alpha|} \right) \frac{a_s}{9(2\kappa_s)^{2/3}\hat\xi^{1/3}} \frac{KM}{g_s\mathcal{V}^{2/3}}, \label{lambda3} \\
\lambda_4 &\equiv \text{max}\left(\frac{1}{|\alpha-1|},\frac{1}{|\frac{9}{4}-\alpha|} \right) \frac{\alpha}{2(g_sM)^2} \left(1+ \frac{3}{\sqrt{9-\frac{64\pi c^\prime c^{\prime\prime}}{g_sM^2}}}\right), \\
 \lambda_5 &\equiv \text{max}\left(\frac{32}{|\alpha-1|},\frac{88}{|\frac{9}{4}-\alpha|} \right)\frac{a_s}{27(2\kappa_s)^{2/3}\hat\xi^{1/3}} \frac{|W_0|^2}{\mathcal{V}^{2/3}}. \label{lambda0}
\end{align}
The meaning of the various control parameters is as follows: $\lambda_1$ controls corrections arising from a small-$g_s$ expansion of the solution. Note that these corrections are not due to explicit loop corrections to the LVS potential but occur already when the leading LVS potential is minimized \cite{Junghans:2022exo}.

The parameter $\lambda_2$ controls a novel type of loop correction in the presence of the conifold modulus. As shown in \cite{Junghans:2022exo}, it arises because the well-known ``KK-loop'' corrections to the K\"ahler potential \cite{Berg:2005ja, Berg:2007wt, Cicoli:2007xp, Junghans:2014zla} mix with terms depending on the conifold modulus upon computing the $F$-term scalar potential.

The parameter $\lambda_3$ is related to a warping correction to the BBHL term in the K\"ahler potential.
Its existence can be inferred from the curvature correction $\propto \int_X \e^{2A-3\phi/2} c_3(X)$ that was already mentioned below \eqref{logxi}.
In particular, for non-constant warp factor, the integral can be expanded in $1/\mathcal{V}$ and yields a correction to the BBHL term at the relative order $KM/\mathcal{V}^{2/3}$ \cite{Junghans:2022exo}.\footnote{See also earlier results in \cite{Carta:2019rhx, Gao:2020xqh} on warping from conifold fluxes.}

$\lambda_4$ is due to a curvature correction to the anti-brane tension, which arises when the anti-brane is placed in a conifold background. This was shown in \cite{Junghans:2022exo} by evaluating known $R^2$ curvature corrections to the DBI action \cite{Bachas:1999um} on the Klebanov-Strassler solution \cite{Klebanov:2000hb}. Finally, $\lambda_5$ is related to higher $F$-terms that descend from curvature corrections involving powers of $G_3$ flux \cite{Cicoli:2013swa, Ciupke:2015msa}.

All of the above corrections modify the $F$-term scalar potential and thus correct the vacuum energy and the moduli masses. Demanding that these corrections can be self-consistently neglected then yields the above conditions for the $\lambda_i$ parameters.
For a more detailed discussion and derivation of these corrections we refer to \cite{Junghans:2022exo}.\footnote{
Note that \cite{Gao:2022uop} recently argued for additional loop corrections with a more general K\"ahler-moduli dependence than considered in \cite{Berg:2007wt, Cicoli:2007xp, Junghans:2022exo}, which might lead to further constraints.}

The above conditions are necessary but not sufficient for a controlled vacuum. Indeed, one also needs to make sure that corrections to the moduli vevs are small. General expressions for the corrected moduli vevs were derived in \cite{Junghans:2022exo}. For the analysis of this paper, it will be sufficient to focus on one type of such corrections, which is rather constraining. In particular, loop corrections shift the exponent of \eqref{tb0} by a term $-\mathcal{C}_s^\text{KK} a_s g_s/3\kappa_s$, which needs to be small in order to retain control over the dS uplift.\footnote{Here $\mathcal{C}_s^\text{KK}$ is a model-dependent numerical coefficient which we again assume to be $\mathcal{O}(1)$.} It is convenient to write this condition in terms of a $\lambda_i$ parameter analogously to the other conditions, i.e.,
\begin{equation}
\lambda_6 \equiv \frac{a_s g_s}{3\kappa_s}, \label{lambda6}
\end{equation}
and demand that \eqref{lc} holds for all $i=1,\ldots,6$.

It is furthermore convenient to define
\begin{equation}
\lambda \equiv \text{max}\left(\lambda_1,\lambda_2,\lambda_3,\lambda_4,\lambda_5,\lambda_6\right). \label{la}
\end{equation}
We can then equivalently write \eqref{lc} as
\begin{equation}
\lambda\ll 1. \label{lc2}
\end{equation}
As shown in \cite{Junghans:2022exo}, it is impossible to satisfy \eqref{lc2} anywhere in the parameter space of the model \eqref{crino}.
Indeed, assuming \eqref{crino} together with the dS condition \eqref{range},
one can prove that
\begin{equation}
\lambda > 0.605 \label{hl}
\end{equation}
for all values of $g_s$, $W_0$, $K$ and $M$.
In this paper, we will generalize this to a bound of the form
\begin{equation}
\lambda > \lambda_\text{min}\!\left(Q_3,\hat\xi, \kappa_s, \chi_s,a_s\right)
\end{equation}
applicable to arbitrary topologies and orientifold/brane data.

\end{itemize}

\subsection{Parametric tadpole constraint}
\label{sec:ptc0}

As stated before, the bound \eqref{hl} holds under the assumption that $Q_3$, $\hat\xi$, $\kappa_s$, $\chi_s$ and $a_s$ take the values \eqref{crino} of the model of \cite{Crino:2020qwk}. A natural question is whether one can improve control
by changing these parameters.
It was argued in \cite{Gao:2022fdi} that this is indeed possible in models with a sufficiently large $Q_3$.
In particular, \cite{Gao:2022fdi} focussed on a specific combination of $\lambda_3$ and $\lambda_5$ (denoted by $1/c_N$ and $1/c_{W_0}$ there) and pointed out that it becomes small for sufficiently large $N\equiv KM$ (which satisfies $N\le Q_3$ due to the tadpole condition).

Let us now reproduce this argument in our framework. We note in advance that \cite{Gao:2022fdi} makes several assumptions on the anti-brane uplift term that differ from our approach here and in \cite{Crino:2020qwk, Junghans:2022exo} and furthermore defines the requirements for control differently than we do (see App.~\ref{sec:ptc} for details). The equation we will derive below therefore only matches with the one derived in \cite{Gao:2022fdi} up to the value of a coefficient.
This is not relevant for the arguments in this section but affects the numerical strength of the resulting constraint.

In our conventions, the basic claim of \cite{Gao:2022fdi} follows from the relation
\begin{align}
\lambda_3\lambda_5^{2/3} &= \gamma N \e^{-N/N_*}, 
\label{ptc0}
\end{align}
where
\begin{equation}
N_* = \frac{9g_sM^2}{16\pi}, \quad \gamma=\text{max}\left(\frac{20}{|\alpha-1|},\frac{55}{|\frac{9}{4}-\alpha|} \right)^{5/3} \frac{16 a_s^{7/3}}{729(2\kappa_s)^{14/9}\hat\xi^{7/9}} \frac{q_0^{2/3}}{g_s^{4/3}\alpha^{2/3}}
\e^{-\frac{2}{3}+\frac{2}{3}\sqrt{1-\frac{64\pi c^{\prime}c^{\prime\prime}}{9g_sM^2}}}
\end{equation}
up to subleading terms in $g_s$, as can be verified using \eqref{ts0}, \eqref{zeta0}, \eqref{alpha}, \eqref{lambda3} and \eqref{lambda0}.
We thus see that $\lambda_3\lambda_5^{2/3}$ becomes small if $N$ is chosen large enough. Since the tadpole condition implies $N\le Q_3$, this requires models with a sufficiently large tadpole $Q_3$.

For completeness, let us also derive a slightly different version of \eqref{ptc0}, which is of the same form as the main result of \cite{Gao:2022fdi}.
To this end, we use the relation
\begin{equation}
\frac{8\pi N}{5M^2}=\frac{a_s\hat\xi^{2/3}}{(2\kappa_s)^{2/3}} - \frac{a_s\chi_s}{24} +\mathcal{O}(g_s). \label{xx0}
\end{equation}
This must hold in any dS vacuum because
\eqref{alpha} implies $ -\frac{4}{5}\ln\zeta\simeq\ln\mathcal{V}$ and therefore $\frac{8}{5}\frac{\pi N}{g_sM^2}=a_s\tau_s-\frac{a_s\chi_s}{24g_s} +\mathcal{O}(g_s^0)$
when $\alpha$ is in the dS range \eqref{range} and the other parameters in \eqref{alpha} are not exponentially small or large.\footnote{Similar arguments relating the conifold fluxes to the vev of the K\"ahler modulus were used in the context of KKLT in \cite{Carta:2019rhx, Gao:2020xqh}.}

Using \eqref{xx0}, we can rewrite \eqref{ptc0} as
\begin{align}
\lambda_3\lambda_5^{2/3} &= \frac{3^{7/3}\gamma^{\prime}}{7^{7/3} N_*^{2}} N^{7/3} \e^{-N/N_*}, \qquad \gamma^\prime= \gamma \left[ \frac{3^{7/4}}{7^{7/4}} \frac{10a_s}{9g_s\sqrt{N_*}}\left(\frac{\hat\xi^{2/3}}{(2\kappa_s)^{2/3}} - \frac{\chi_s}{24}\right)\right]^{-4/3}. \label{ptc1}
\end{align}
Solving this for $N$ yields
\begin{equation}
\frac{N}{N_*}= -\frac{7}{3}\mathcal{W}_{-1}\left( -\frac{\lambda_3^{3/7}\lambda_5^{2/7}}{\gamma^{\prime 3/7} N_*^{1/7}}\right), \label{ptc2}
\end{equation}
where $\mathcal{W}_{-1}$ is the $-1$ branch of the Lambert W-function. We finally expand the right-hand side assuming a small argument and thus arrive at
\begin{equation}
\frac{N}{N_*}=\frac{1}{3}\ln N_* -\ln\lambda_3-\frac{2}{3}\ln\lambda_5 + \ln\gamma^\prime + \mathcal{O}(\ln\ln). \label{ptc3}
\end{equation}
This equation was called the ``LVS parametric tadpole constraint'' in \cite{Gao:2022fdi} and is referred to as PTC in the following.
As stated before, there is a disagreement in the coefficient $\gamma^\prime$ as defined here and the corresponding coefficient in the version of the PTC derived in \cite{Gao:2022fdi}. This can be traced back to different definitions of the anti-brane uplift and the control parameters (see App.~\ref{sec:ptc}).
The precise value of $\gamma^\prime$ is not relevant for the qualitative point we wish to make in this section.

In any case, we again see from \eqref{ptc3} that $\lambda_3\lambda_5^{2/3}$ becomes small if $N\le Q_3$ is sufficiently large.
Based on this observation, \cite{Gao:2022fdi} argued that the PTC
is the main obstacle in constructing self-consistent LVS dS vacua.

However, this is actually not the case.
Indeed,
the PTC can be
deceiving, as
it is only sensitive to the product $\lambda_3\lambda_5^{2/3}$ but not to $\lambda_3$ and $\lambda_5$ individually and not to the remaining $\lambda_i$ parameters either.
One of the main points of this paper is that satisfying the full set of conditions $\lambda_i\ll 1$ (as in \cite{Junghans:2022exo}) is much harder than satisfying the proxy $\lambda_3\lambda_5^{2/3}\ll 1$ and yields much stronger constraints on the LVS.

As a concrete example, consider the model of \cite{Crino:2020qwk} specified by \eqref{crino}. We furthermore choose $\alpha=\frac{4}{3}$, $K=7$, $M=21$, $g_s=\frac{64\pi c^\prime c^{\prime\prime}}{9M^2}\approx 0.105$, which implies $g_sM^2\approx 46.1$ and $N=147$ and is thus consistent with \eqref{gsm} and the tadpole $Q_3=149$ in this model. Substituting these parameter choices into \eqref{ptc1}, we find\footnote{As also noted in \cite{Gao:2022fdi}, the approximation of discarding the $\mathcal{O}(\ln\ln)$ terms in \eqref{ptc3} yields a large error when exponentiating so that one should use the exact equation \eqref{ptc2} or, equivalently, \eqref{ptc1} instead.}
\begin{equation}
\lambda_3\lambda_5^{2/3}\approx 5.61\cdot 10^{-4}, \label{ptcex}
\end{equation}
which is satisfied, e.g., for $\lambda_3=\lambda_5\approx  1.12\cdot 10^{-2} \ll 1$.

We thus see that the PTC is by itself not an obstruction to achieving excellent control in this model. On the other hand, a controlled dS vacuum would be inconsistent with \eqref{hl} so that we must have missed something.
Indeed, evaluating \eqref{lambda3}, \eqref{lambda0} individually, we find $\lambda_3\approx 1.61\cdot 10^7$, $\lambda_5\approx 1.38\cdot 10^{-16}$ and therefore huge warping corrections, showing that the example is completely out of control.\footnote{
One can furthermore check that this example has
a tiny, fine-tuned $W_0\approx 1.50\cdot 10^{-10}$ and a tiny volume $\mathcal{V}\approx 6.47\cdot 10^{-5}$, again indicating a loss of control. One could attempt to exclude such pathological examples and thus strengthen the constraint on the tadpole by imposing $W_0=\mathcal{O}(\sqrt{N})$ or $\mathcal{V}\gg 1$ in conjunction with the PTC. However, it is not obvious how to turn these separate conditions into an explicit, quantitative bound on $N$, as the PTC itself is independent of both $W_0$ and $\mathcal{V}$.} One can check \cite{Junghans:2022exo} that all other solutions in this model also have at least one large $\lambda_i$ parameter (for $i=1,\ldots,6$) or are outside of the dS range \eqref{range}.
The PTC is oblivious to this problem, as it only sees one particular product of the $\lambda_i$ parameters, which is small here.
In other words, the fact that the PTC is consistent with $\lambda_i\approx 0.01$ for $N= 147$ in this model does not imply the existence of dS solutions with these parameter values.

This is explained as follows. First, the PTC depends on the parameter $N_*\sim g_sM^2$ whose relation to the $\lambda_i$ parameters is not obvious, so that the requirements for control are not entirely manifest.
While it was assumed in \cite{Gao:2022fdi} that $g_sM^2\approx 46.1$ (as in the above example) is sufficient for control, one could in principle also consider larger values of $g_sM^2$ in the PTC, which would increase the required $N$ and thus the bound on $Q_3$. However, without a computation that quantitatively connects the value of $g_sM^2$ to the $\lambda_i$ parameters, it is not clear which value of $g_sM^2$ would be required to achieve a certain degree of control.
A second important point is that we only have four adjustable parameters $g_s$, $W_0$, $K$, $M$ in a given model, which is not enough freedom to fix every $\lambda_i$ parameter to a desired value and at the same time satisfy the dS condition \eqref{range}. Therefore, a dS solution may simply not exist for a given choice of the $\lambda_i$ parameters, even though this choice may be consistent with the PTC. As a consequence, the actual requirement for $Q_3$ is much stronger in typical models than predicted by the PTC.

For the above reasons, imposing all of the conditions $\lambda_i\ll 1$ is crucial in order to appreciate how strongly constrained the LVS is.
In the remainder of this work, we will derive a bound
based on this requirement.
We will also see that the basic idea of \cite{Gao:2022fdi} remains correct: large tadpoles indeed improve control.
However, our constraints differ significantly on a quantitative level. For example, assuming that $\hat\xi$, $\kappa_s$, $\chi_s$, $a_s$ take the values in \eqref{crino}, our results in Section \ref{sec:bound} imply that $\lambda_i\le 0.1$ requires $Q_3\gtrsim 3.59\cdot 10^3$ and $\lambda_i\le 0.01$ requires $Q_3\gtrsim 3.51\cdot 10^6$, whereas we saw above that much smaller tadpoles $Q_3\sim 150$ seem perfectly fine from the point of view of the PTC.

\section{Derivation of the bound}
\label{sec:bound}

In this section, we derive a bound on the $\lambda_i$ parameters valid for general LVS models of the type reviewed in Section \ref{sec:review1}.
The goal is to eliminate the dependence on the moduli vevs and the tunable parameters $g_s$, $W_0$, $\alpha$, $K$ and $M$ and thus arrive at a bound that only depends on $Q_3$, $\hat\xi$, $\kappa_s$, $\chi_s$ and $a_s$.
To this end, it is useful to consider the following combinations of the $\lambda_i$ parameters:
\begin{equation}
\bar\lambda_1\equiv \lambda_1, \qquad \bar\lambda_2\equiv \lambda_2^{8/9}\lambda_5^{1/9}, \qquad \bar\lambda_3\equiv \lambda_3^{2/3}\lambda_5^{1/3}, \qquad \bar\lambda_4\equiv \lambda_4, \qquad \bar\lambda_5\equiv\lambda_5, \qquad \bar\lambda_6\equiv\lambda_6. \label{barl}
\end{equation}
We further define
\begin{equation}
\bar\lambda \equiv \text{max}\left(\bar\lambda_1,\bar\lambda_2,\bar\lambda_3,\bar\lambda_4,\bar\lambda_5,\bar\lambda_6\right) \label{bb}
\end{equation}
in analogy with \eqref{la}. It then follows from \eqref{barl} that
\begin{equation}
\lambda\ge\bar\lambda. \label{bb1}
\end{equation}
It is also convenient to work with
\begin{equation}
N\equiv KM,\qquad k\equiv\frac{K}{M} \label{nk}
\end{equation}
instead of $K$, $M$ and to abbreviate
\begin{equation}
P\equiv \hat\xi^{2/3}, \qquad R\equiv \hat\xi^{2/3}-\frac{(2\kappa_s)^{2/3}\chi_s}{24},\qquad S\equiv\frac{(2\kappa_s)^{2/3}}{a_s}. \label{nprs}
\end{equation}
Using this notation, \eqref{xx0} becomes
\begin{equation}
\frac{8\pi k}{5}=\frac{R}{S} +\mathcal{O}(g_s). \label{xx}
\end{equation}
Note that \eqref{xx} is a non-trivial constraint since the possibilities to adjust $k$ are limited by the requirements of flux quantization and tadpole cancellation. We will come back to this point in Section \ref{sec:disc2} and assume for now that \eqref{xx} is satisfied.

Using the above together with \eqref{tb0}, \eqref{ts0}, \eqref{alpha}, \eqref{q0} in the $\bar\lambda_i$ parameters (defined by \eqref{barl}, \eqref{lambda00}--\eqref{lambda6}), we find
\begin{align}
\bar\lambda_1 &= \text{max}\left(\frac{10+\alpha+4\alpha^2}{30|\alpha-1|}, \frac{135-16\alpha+16\alpha^2}{120|\frac{9}{4}-\alpha|} \right) \frac{g_sS}{P}, \label{lambda10} \\
\bar\lambda_2 &= \text{max}\left(\frac{\left|7r-5\right|}{|\alpha-1|},\frac{\left|10r-5\right|}{|\frac{9}{4}-\alpha|} \right)^{8/9} \text{max}\left(\frac{32}{|\alpha-1|},\frac{88}{|\frac{9}{4}-\alpha|} \right)^{1/9} \nl \times
\frac{2^{5/9}3^{2/9}5^{2/3}\pi^{16/9} 512 c^{\prime 4/3} g_s^3 N^{8/9} \alpha^{4/9}}{675R^{8/9}P^{5/6} S^{7/9} \left(1-r\right)^{4/9}r^{8/9}} \exp\left({-\frac{10R}{9 g_sS} - \frac{10}{27} - \frac{8\alpha}{27} }\right), \label{lambda20} \\
\bar\lambda_3 &= \text{max}\left(\frac{20}{|\alpha-1|},\frac{55}{|\frac{9}{4}-\alpha|} \right)
\frac{5^{2/3} 8 N^{2/3}}{135 g_s^{1/3}P^{5/6}S^{5/3}} \exp\left({-\frac{2R}{3 g_sS} - \frac{2}{9}-\frac{8\alpha}{45}}\right), \label{lambda30} \\
\bar\lambda_4 &= \text{max}\left(\frac{1}{|\alpha-1|},\frac{1}{|\frac{9}{4}-\alpha|} \right) \frac{5\alpha R}{16\pi g_s^2SN} \left(1+ \frac{1}{r}\right), \label{lambda40} \\
\bar\lambda_5 &= \text{max}\left(\frac{32}{|\alpha-1|},\frac{88}{|\frac{9}{4}-\alpha|} \right) \frac{3^{1/3}4 g_s^{1/3}|W_0|^{4/3}}{81 P^{5/6}S^{5/3} } \exp\left({-\frac{2R}{3g_sS} - \frac{2}{9} - \frac{8\alpha}{45}}\right), \label{lambda50} \\
\bar\lambda_6 &= \frac{2g_s}{3\sqrt{a_s}S^{3/2}}, \label{lambda60}
\end{align}
where $r=\sqrt{1-\frac{40 c^\prime c^{\prime\prime}R}{9g_sSN}}$, $c^\prime =1.18$, $c^{\prime\prime}=1.75$ and we ignored subleading terms in $g_s$.\footnote{In particular, we used that $\sqrt{1-\frac{64\pi c^\prime c^{\prime\prime}}{9g_sM^2}}\approx r$ according to \eqref{nk}, \eqref{xx}. The approximation of dropping the $g_s$ corrections to \eqref{xx} in the square root is justified in the large-$N$ regime, which is the regime considered further below.}
Note that we eliminated the dependence on the moduli vevs and on $k$ in all $\bar\lambda_i$ parameters.

We further observe that all $\bar\lambda_i$ parameters except for $\bar\lambda_5$ are independent of $W_0$.
This means that once we find a solution with $\bar\lambda_1,\bar\lambda_2,\bar\lambda_3,\bar\lambda_4,\bar\lambda_6 \le c$ for some small number $c$, we can always ensure that $\bar\lambda_5\le c$ as well by an appropriate choice of $W_0$ that leaves the other $\bar\lambda_i$ parameters invariant.
In particular, the requirement of small $\bar\lambda_5$ does not constrain the topological parameters we are interested in. We can therefore ignore $\bar\lambda_5$ in the following and focus on the constraints from $\bar\lambda_1$ to $\bar\lambda_4$ and $\bar\lambda_6$.

The next step is to eliminate the $\alpha$ dependence in the remaining $\bar\lambda_i$ parameters. To this end, we follow the same strategy as in \cite{Junghans:2022exo}. In particular, we can bound each $\bar\lambda_i$ from below using that $\bar\lambda_i\ge \hat\lambda_i\equiv \bar\lambda_i(\hat\alpha_i)$, where we denote by $\hat\alpha_i$ the value of $\alpha\in]1,\frac{9}{4}[$ for which $\bar\lambda_i$ is minimized.
Defining further
\begin{equation}
\hat\lambda \equiv \text{max}\left(\hat\lambda_1,\hat\lambda_2,\hat\lambda_3,\hat\lambda_4, \hat\lambda_6\right),
\end{equation}
we have
\begin{equation}
\bar\lambda\ge\hat\lambda. \label{bb2}
\end{equation}
Minimizing the $\bar\lambda_i$ parameters with respect to $\alpha$, we find\footnote{The expression for $\hat\alpha_2$ is correct for $r\ge \frac{40}{53}$. This is sufficient for our purpose, as we will be interested in the regime $r\approx 1$ below.
}
\begin{align}
\hat \alpha_1 &= \frac{(3311+495\sqrt{82})^{1/3}}{12}-\frac{209}{12(3311+495\sqrt{82})^{1/3}}+\frac{2}{3} \approx 1.44, \notag \\
\hat \alpha_2 &= \frac{9\left|7r-5\right|+20\left|2r-1\right|}{4\left|7r-5\right|+20\left|2r-1\right|}, \qquad
\hat \alpha_3 = \frac{4}{3}, \qquad \hat \alpha_4 = \frac{13}{8} \label{hatalpha}
\end{align}
and therefore
\begin{align}
\hat\lambda_1 &= \Big((991-69\sqrt{82})(3311 + 495\sqrt{82})^{2/3}-(2299- 1254\sqrt{82})(3311 + 495\sqrt{82})^{1/3}\nl +210463  \Big) \frac{g_s S}{357390P}\approx 1.49\frac{g_s S}{P}, \label{hat1} \\
\hat\lambda_2 &= \frac{3^{2/9}\pi^{16/9} 2048 c^{\prime 4/3} g_s^3 N^{8/9} \left[\left(|7r - 5| + 5|2r - 1|\right)\left(9|7r - 5| + 20|2r - 1|\right)\right]^{4/9}}{5^{1/3}675R^{8/9}P^{5/6} S^{7/9} \left(1-r\right)^{4/9}r^{8/9}}
\nl\times \text{max}\left(\frac{11(|7r - 5| + 5|2r - 1|)}{5|2r - 1|}, \frac{4(|7r - 5| + 5|2r - 1|)}{|7r - 5|}\right)^{1/9} \nl\times 
\exp\left({-\frac{10R}{9 g_sS} - \frac{10}{27} -\frac{18|7r - 5| + 40|2r - 1|}{27|7r - 5| + 135|2r - 1|} }\right)
, \label{hat2} \\
\hat\lambda_3 &= \frac{5^{2/3} 32 N^{2/3}}{9 g_s^{1/3} P^{5/6}S^{5/3}} \exp\left({- \frac{2R}{3g_sS} -\frac{62}{135}}\right), \label{hat3}\\
\hat\lambda_4 &= \frac{13R}{16\pi g_s^2SN}\left(1 + \frac{1}{r}\right), \label{hat4} \\
\hat\lambda_6 &= \frac{2g_s}{3\sqrt{a_s}S^{3/2}}. \label{hat6}
\end{align}

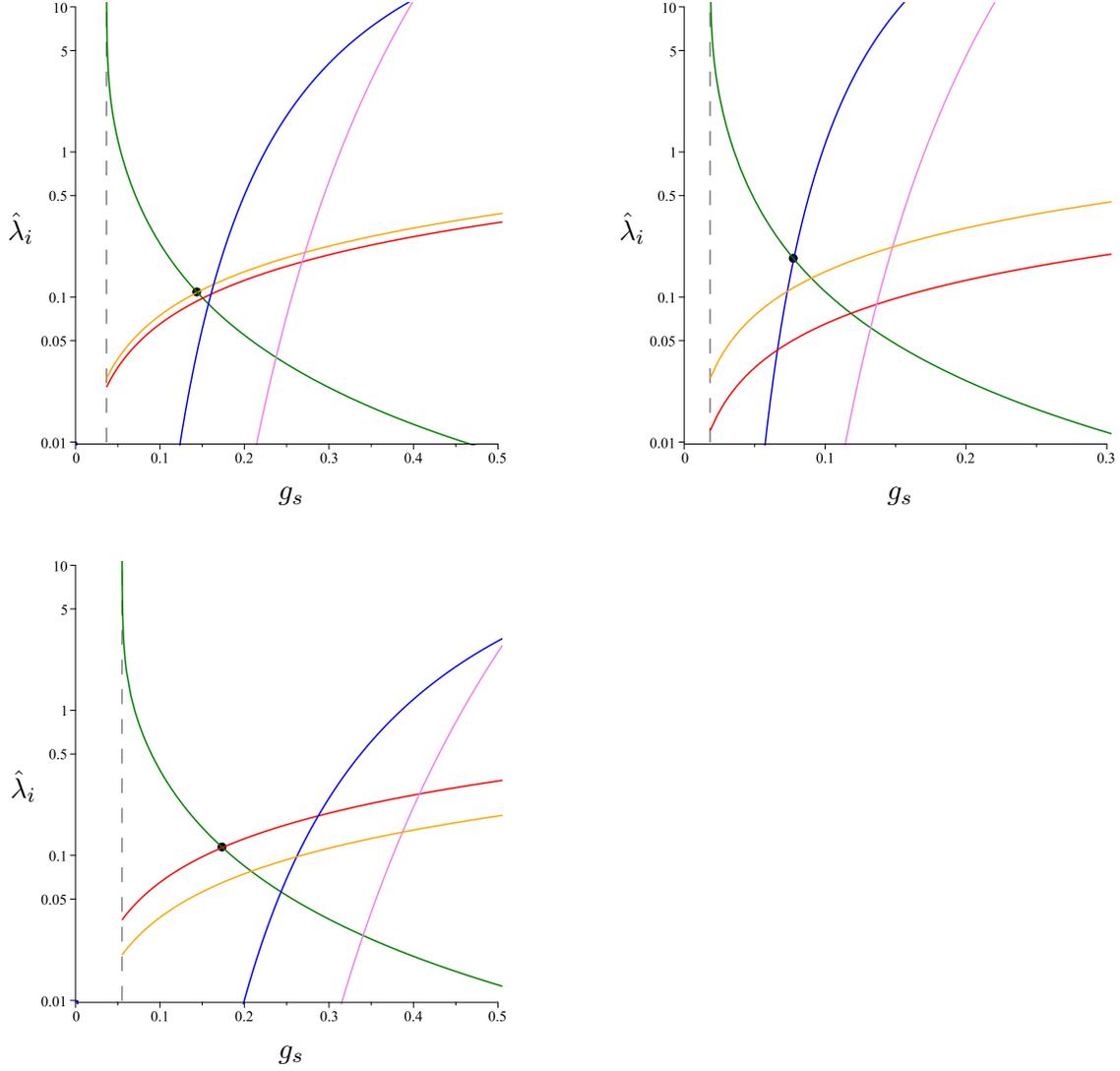
\begin{figure}[p]
\includegraphics[trim = 0mm 0mm 0mm 0mm, clip, width=0.43\linewidth]{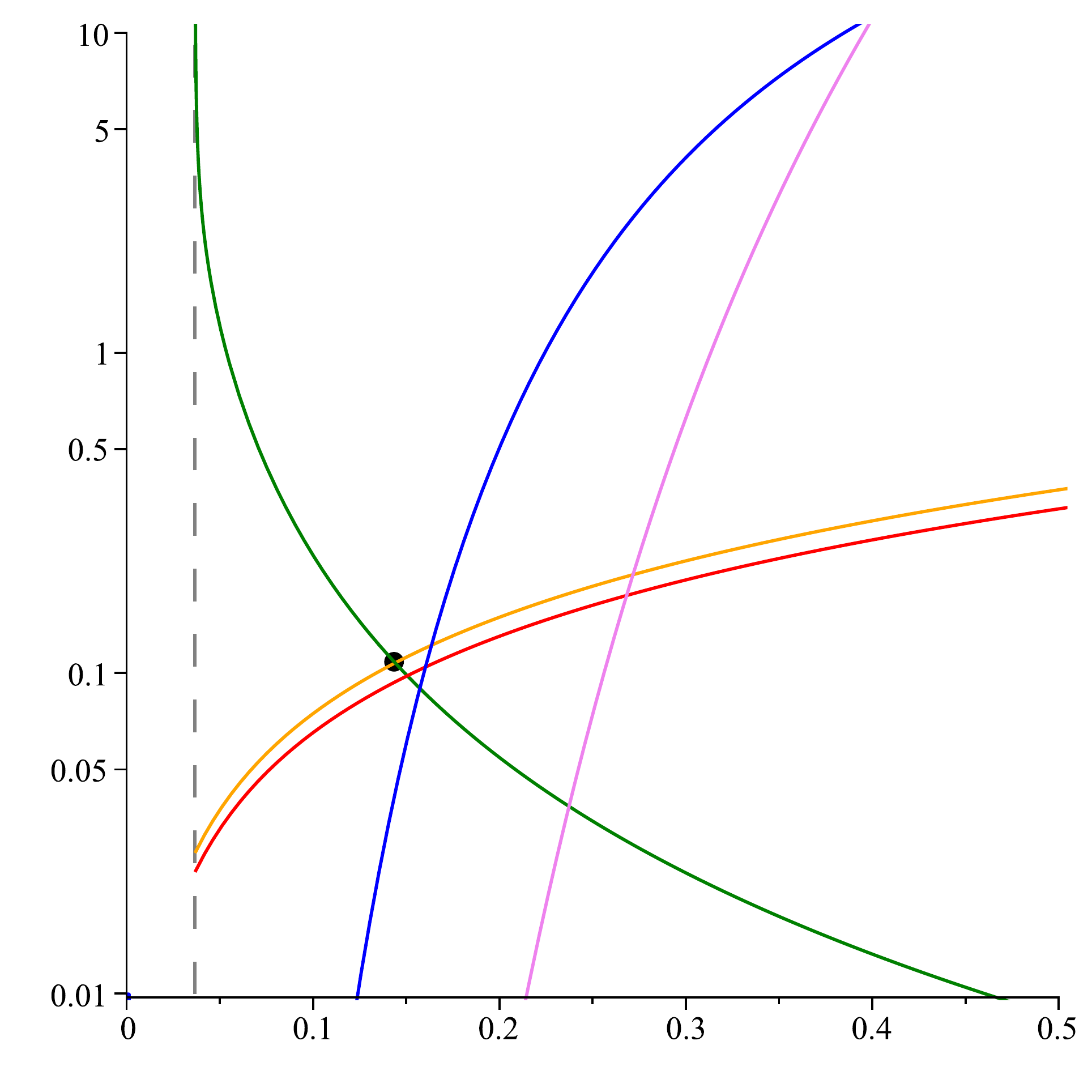}\qquad\qquad\includegraphics[trim = 0mm 0mm 0mm 0mm, clip, width=0.43\linewidth]{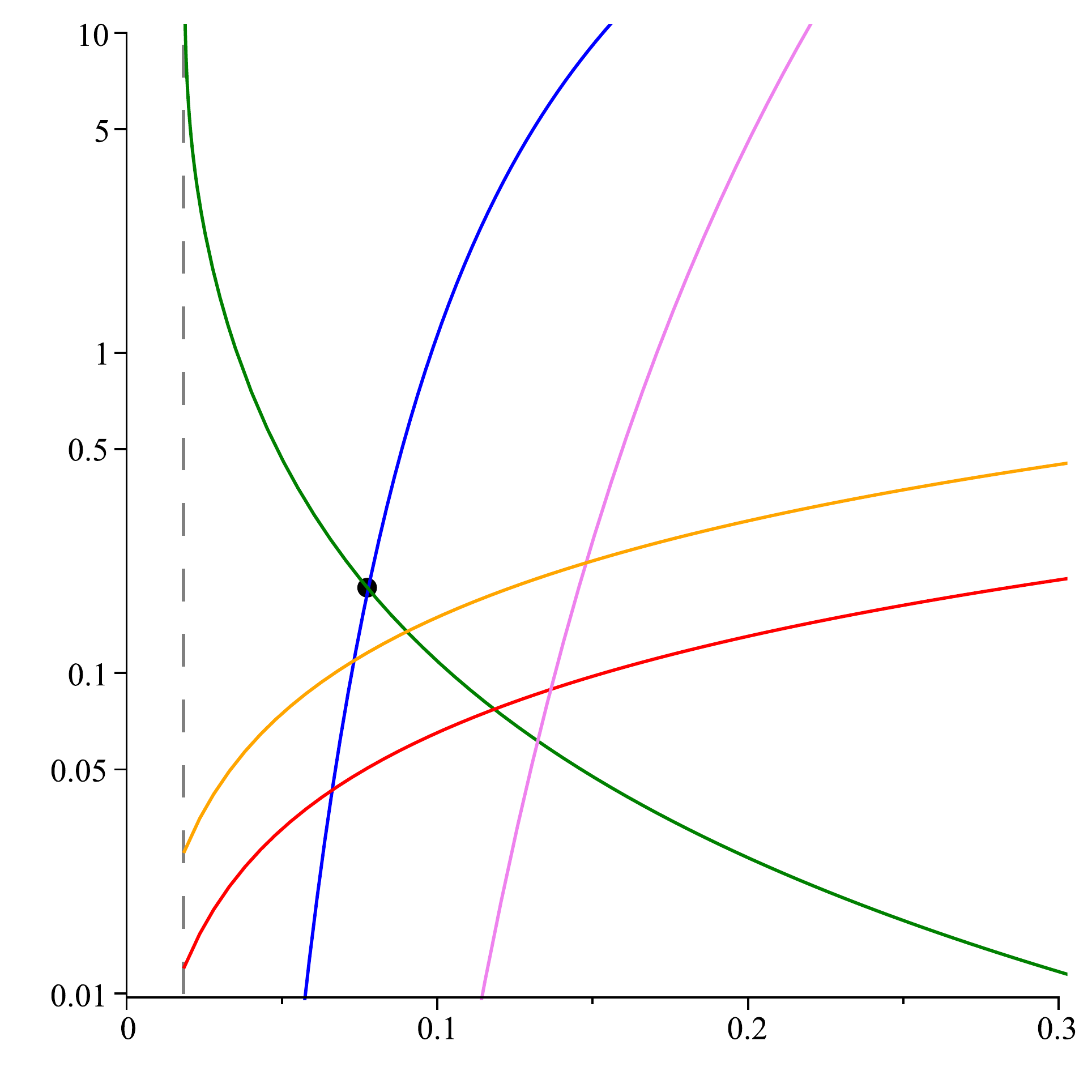}
\put(-428,95){$\hat\lambda_i$}
\put(-324,-5){$g_s$}
\put(-193,95){$\hat\lambda_i$}
\put(-90,-5){$g_s$}
\vspace{1.5em}

\includegraphics[trim = 0mm 0mm 0mm 0mm, clip, width=0.43\linewidth]{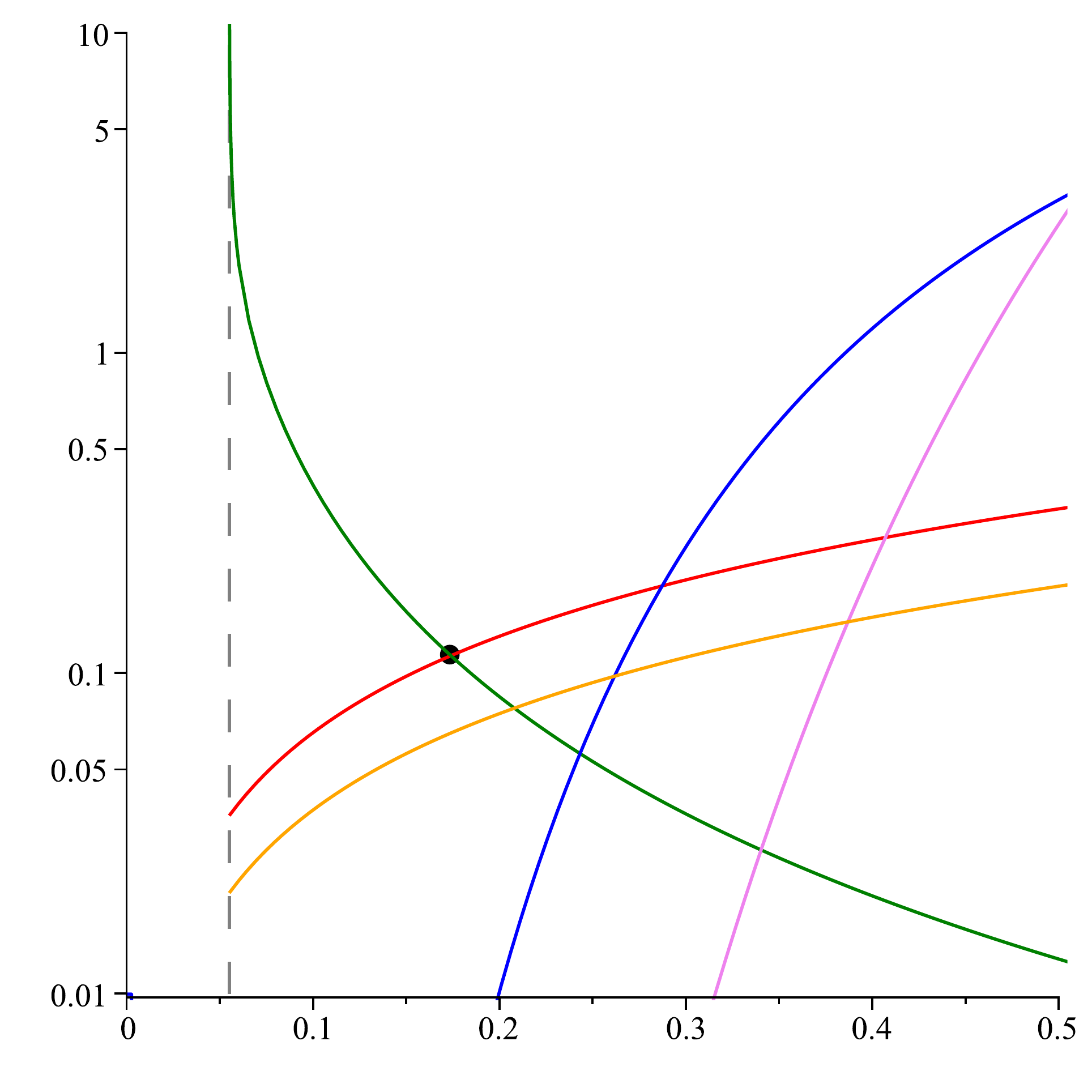}
\put(-193,95){$\hat\lambda_i$}
\put(-90,-5){$g_s$}
\vspace{1em}

\caption{$\hat\lambda_1$ (orange), $\hat\lambda_2$ (violet), $\hat\lambda_3$ (blue), $\hat\lambda_4$ (green) and $\hat\lambda_6$ (red) as functions of $g_s$ for $N=500$, $S=1$, $a_s=\frac{\pi}{3}$ and $P=R=2$ (top left), $P=R=1$ (top right), $P=4$, $R=3$ (bottom left). The grey dashed lines are due to \eqref{gsm}, which together with \eqref{nk}, \eqref{xx} implies $g_s\ge 40c^\prime c^{\prime\prime}R/9SN$.
The minimal $\hat\lambda$ in each plot is located at the intersection of $\hat\lambda_4$ with one of the other $\hat\lambda_i$ parameters (marked by the black dots). Note that the intersection of $\hat\lambda_4$ and $\hat\lambda_2$ is not relevant for the bound unless some of the other corrections are absent.\\}
\label{l-plot}
\end{figure}

The final step is to minimize $\hat\lambda$ with respect to $g_s$
in order to get a bound on the maximally possible control in a given model.
Hence, we need to determine the value of $g_s$ for which the largest of the expressions \eqref{hat1}--\eqref{hat6} is as small as possible. As shown in Fig.~\ref{l-plot}, this happens at the value of $g_s$ where $\hat\lambda_4$ intersects with either $\hat\lambda_1$, $\hat\lambda_2$, $\hat\lambda_3$ or $\hat\lambda_6$, where out of these four possibilities we have to choose the one where $\hat\lambda_4$ is the largest. Let us denote the four values of $\hat\lambda_4$ at the intersections by $\hat\lambda^{(14)}$, $\hat\lambda^{(24)}$, $\hat\lambda^{(34)}$ and $\hat\lambda^{(64)}$. It is evident from Fig.~\ref{l-plot} that we then have
\begin{equation}
\hat \lambda \ge \text{max}\left( \hat\lambda^{(14)} , \hat\lambda^{(24)}, \hat\lambda^{(34)}, \hat\lambda^{(64)} \right). \label{bb3}
\end{equation}

We consider the intersection of $\hat\lambda_1$ and $\hat\lambda_4$ first. The intersection locus is described by the equation $\hat\lambda_1=\hat\lambda_4$, i.e.,
\begin{equation}
\frac{1.49 g_s S}{P} = \frac{13R\left(1 + \frac{1}{\sqrt{1- \frac{40 c^\prime c^{\prime\prime}R}{9g_sSN}}}\right)}{16\pi g_s^2SN}. \label{ie}
\end{equation}
We can solve this for $g_s$ in an expansion in $1/N$. This yields the solution
\begin{equation}
g_s^{(14)}=\frac{0.702 P^{1/3}R^{1/3}}{S^{2/3}N^{1/3}} + \frac{0.765 R}{SN} + \frac{5.83 R^{5/3}}{P^{1/3}S^{4/3}N^{5/3}} + \mathcal{O}\left(\frac{R^{7/3}}{P^{2/3}S^{5/3}N^{7/3}}\right). \label{gs14}
\end{equation}
Substituting $g_s=g_s^{(14)}$ into $\hat\lambda_1$ or $\hat\lambda_4$, we get
\begin{equation}
\hat\lambda^{(14)}\equiv \hat\lambda_1(g_s^{(14)})=\hat\lambda_4(g_s^{(14)}) = \frac{1.05 S^{1/3}R^{1/3}}{P^{2/3}N^{1/3}} + \frac{1.14 R}{PN} + \frac{8.71 R^{5/3}}{P^{4/3}S^{1/3}N^{5/3}} + \mathcal{O}\left(\frac{R^{7/3}}{P^{5/3}S^{2/3}N^{7/3}}\right). \label{hl2}
\end{equation}

We now consider the intersection of $\hat\lambda_2$ and $\hat\lambda_4$. We thus need to solve the equation $\hat\lambda_2=\hat\lambda_4$.
Solving this for $g_s$ in a $1/N$ expansion yields\footnote{Evaluating the factor $\text{max}(a,b)^{1/9}$ in \eqref{hat2} is non-trivial, as $a$ and $b$ are functions of $g_s$, $N$, $R$, $S$. Here, we assume $\text{max}(a,b)=b$, as $b>a$ in the relevant regime of moderately large $N$, where $r$ is close to (but not exactly) 1. Asymptotically in the large-$N$ limit (i.e., at $r=1$), we instead have $a>b$. However, the difference is negligible, $(a/b)^{1/9}\approx 1.01$, so that we refrain from considering this case separately.}
\begin{align}
g_s^{(24)}&= \frac{10R}{49S\, \mathcal{W}^{(24)}} + \frac{315880R c^\prime c^{\prime\prime}}{410571 S \left(\mathcal{W}^{(24)} +1\right)N}
\nl + \frac{144590330 R c^{\prime 2} c^{\prime\prime 2} \mathcal{W}^{(24)}\left((\mathcal{W}^{(24)})^2+2\mathcal{W}^{(24)}+\frac{4460547447}{4959448319}\right)}{10029663S \left(\mathcal{W}^{(24)} +1\right)^3N^2}
 + \mathcal{O}\left(\frac{R}{SN^3}\right), \label{gs24}
\end{align}
where $\mathcal{W}^{(24)}\equiv \mathcal{W}_0\!\left(z^{(24)}\right)$ is the principal branch of the Lambert W-function with argument
\begin{align}
z^{(24)}&=\frac{8\cdot 2^{25/49}3^{32/49}5^{24/49}7^{5/49}13^{40/49}19^{4/49}\pi^{25/49}c^{\prime 8/49}\e^{-146/1029}}{1911c^{\prime\prime 4/49}}\frac{N^{3/7}R^{4/7}}{P^{15/98}S^{43/49}}\nl\approx 0.517\frac{N^{3/7}R^{4/7}}{P^{15/98}S^{43/49}}.
\end{align}
Substituting $g_s=g_s^{(24)}$ into $\hat\lambda_2$ or $\hat\lambda_4$, we get
\begin{align}
\hat\lambda^{(24)} &\equiv\hat\lambda_2(g_s^{(24)})=\hat\lambda_4(g_s^{(24)}) = \frac{31213S\, (\mathcal{W}^{(24)})^2}{800\pi NR} + \frac{637c^\prime c^{\prime\prime} S\, (\mathcal{W}^{(24)})^3 \left(45619\mathcal{W}^{(24)}-17557 \right)}{136800\pi N^2R\left(\mathcal{W}^{(24)}+1\right)} \nl + \frac{13c^{\prime 2} c^{\prime\prime 2} S\, (\mathcal{W}^{(24)})^4}{23392800\pi N^3R\left(\mathcal{W}^{(24)}+1\right)^3} \left(6243279483(\mathcal{W}^{(24)})^3+4487902895(\mathcal{W}^{(24)})^2\right.\nl\left.-6760627427\mathcal{W}^{(24)}-4007449095 \right) 
+ \mathcal{O}\left(\frac{S}{RN^4}\right). \label{hl3}
\end{align}

We now consider the intersection of $\hat\lambda_3$ and $\hat\lambda_4$. We thus need to solve the equation $\hat\lambda_3=\hat\lambda_4$.
Solving this for $g_s$ in a $1/N$ expansion yields
\begin{align}
g_s^{(34)}&= \frac{2R}{5S\, \mathcal{W}^{(34)}} + \frac{2R c^\prime c^{\prime\prime}}{3S \left(\mathcal{W}^{(34)} +1\right)N}
+ \frac{125R c^{\prime 2} c^{\prime\prime 2} \mathcal{W}^{(34)}\left((\mathcal{W}^{(34)})^2+2\mathcal{W}^{(34)}+\frac{22}{25}\right)}{27S \left(\mathcal{W}^{(34)} +1\right)^3N^2}
\nl + \mathcal{O}\left(\frac{R}{SN^3}\right), \label{gs34}
\end{align}
where $\mathcal{W}^{(34)}\equiv \mathcal{W}_0\!\left(z^{(34)}\right)$ is the principal branch of the Lambert W-function with argument
\begin{equation}
z^{(34)}=\frac{2^{4/5} 32\pi^{3/5}\e^{-62/225}}{585^{3/5}}\frac{NR^{2/5}}{\sqrt{P}S^{7/5}}\approx 1.84\frac{NR^{2/5}}{\sqrt{P}S^{7/5}}.
\end{equation}
Substituting $g_s=g_s^{(34)}$ into $\hat\lambda_3$ or $\hat\lambda_4$, we get
\begin{align}
\hat\lambda^{(34)} &\equiv\hat\lambda_3(g_s^{(34)})=\hat\lambda_4(g_s^{(34)}) = \frac{325S\, (\mathcal{W}^{(34)})^2}{32\pi NR} + \frac{8125c^\prime c^{\prime\prime} S\, (\mathcal{W}^{(34)})^3 \left(\mathcal{W}^{(34)}-\frac{1}{5} \right)}{288\pi N^2R\left(\mathcal{W}^{(34)}+1\right)} \nl + \frac{8125c^{\prime 2} c^{\prime\prime 2} S\, (\mathcal{W}^{(34)})^4 \left(25(\mathcal{W}^{(34)})^3+35(\mathcal{W}^{(34)})^2+4\mathcal{W}^{(34)}-3 \right)}{864\pi N^3R\left(\mathcal{W}^{(34)}+1\right)^3}
+ \mathcal{O}\left(\frac{S}{RN^4}\right). \label{hl1}
\end{align}

We finally compute the intersection between $\hat\lambda_6$ and $\hat\lambda_4$. This works analogously to the intersection between $\hat\lambda_1$ and $\hat\lambda_4$, except for a different factor multiplying $g_s$ on the left-hand side of \eqref{ie}.
We thus find
\begin{equation}
g_s^{(64)}=\frac{0.919 a_s^{1/6}S^{1/6}R^{1/3}}{N^{1/3}} + \frac{0.765 R}{SN} + \frac{4.46 R^{5/3}}{a_s^{1/6}S^{13/6}N^{5/3}} + \mathcal{O}\left(\frac{R^{7/3}}{a_s^{1/3}S^{10/3}N^{7/3}}\right)
\end{equation}
and
\begin{equation}
\hat\lambda^{(64)} = \frac{0.613 R^{1/3}}{a_s^{1/3}S^{4/3}N^{1/3}} + \frac{0.510 R}{\sqrt{a_s}S^{5/2}N} + \frac{2.97 R^{5/3}}{a_s^{2/3}S^{11/3}N^{5/3}} + \mathcal{O}\left(\frac{R^{7/3}}{a_s^{5/6}S^{29/6}N^{7/3}}\right). \label{hl6}
\end{equation}

The above expressions for $\hat\lambda^{(i 4)}$ are valid in the regime $r\approx 1$, i.e., for $40 c^\prime c^{\prime\prime}R/9g_s^{(i4)} SN \ll 1$.
This yields the conditions
\begin{align}
& \hat\lambda^{(1 4)}: \quad \frac{13.1R^{2/3}}{P^{1/3}S^{1/3}N^{2/3}}\ll 1, && \hat\lambda^{(2 4)}: \quad \frac{45.0\mathcal{W}^{(24)}}{N}\ll 1, \notag \\ & \hat\lambda^{(3 4)}: \quad \frac{22.9\mathcal{W}^{(34)}}{N}\ll 1, && \hat\lambda^{(6 4)}: \quad \frac{9.99R^{2/3}}{a_s^{1/6}S^{7/6}N^{2/3}} \ll 1.
\end{align}
These conditions are satisfied at sufficiently large $N$, which will be the case in the plots shown in the next section.\footnote{For the plots, we keep the first four orders in the large-$N$ expansion of the $\hat\lambda^{(i4)}$ terms. Expanding the square root in $r$ up to this order is a very good approximation for $40 c^\prime c^{\prime\prime}R/9g_s^{(i4)} SN\lesssim 0.8$. Our expressions are therefore reliable even when the left-hand sides of the above conditions are rather close to 1.}

Putting everything together, we conclude from \eqref{bb1}, \eqref{bb2} and \eqref{bb3} that
\begin{align}
\lambda \ge \text{max}\left.\left( \hat\lambda^{(14)} , \hat\lambda^{(24)}, \hat\lambda^{(34)}, \hat\lambda^{(64)} \right)\right|_{N=Q_3}, \label{fb}
\end{align}
where we used that $N\le Q_3$ and that each of the four terms in the bracket (given by \eqref{hl2}, \eqref{hl3}, \eqref{hl1} and \eqref{hl6}) decreases monotonically with $N$ at large $N$.
We have thus computed a lower bound on the amount of control in LVS dS vacua which only depends on $Q_3$, $P$, $R$, $S$ and $a_s$, i.e., on the topology of the CY manifold and the orientifold/brane data. Eqns.~\eqref{hl2}, \eqref{hl3}, \eqref{hl1}, \eqref{hl6} and \eqref{fb} are the main results of this paper.

Note that $\hat\lambda^{(24)}$ is typically smaller than the other terms on the right-hand side of \eqref{fb} and therefore irrelevant for the bound in most cases. However, it might be relevant if there are models where some of the other corrections happen to vanish (or have very small numerical coefficients). We therefore chose to keep $\hat\lambda^{(24)}$ in the above formula for completeness.

\section{Discussion}
\label{sec:disc}

\subsection{Topological constraints on LVS dS vacua}

We can now use the bound \eqref{fb} to constrain LVS models based on the requirement of control.
In particular, we should demand that $\lambda\le c$ for some small number $c$ (for example, $c=0.1$) in order to suppress all unwanted corrections.\footnote{The value of $c$ for which vacua can be considered ``under control'' is debatable since only the parametric scalings but not the numerical coefficients of the various corrections are known in smooth CYs. In particular, the numerical coefficients (which are usually assumed to be $\mathcal{O}(1)$ and set to 1 in our analysis) could a priori
be considerably larger than 1. It is therefore strictly speaking not guaranteed for any finite $c$ that the corrections are suppressed.
Since there are no vacua in the limit $c\to 0$ (which would require unbounded parameters), the best we can do is to make $c$ as small as possible.
}
According to \eqref{fb}, this implies that $\hat\lambda^{(14)}$, $\hat\lambda^{(24)}$, $\hat\lambda^{(34)}$ and $\hat\lambda^{(64)}$ must be smaller than $c$, which in turn rules out regions in the parameter space spanned by $Q_3$, $P$, $R$, $S$ and $a_s$.
Recall that $P$, $R$ and $S$ are defined in terms of the more familar parameters $\hat\chi$, $\kappa_{sss}$, $\chi_s$ and $a_s$ as follows:
\begin{equation}
P= \frac{|\hat\chi|^{2/3}\zeta(3)^{2/3}}{2^{2/3}(2\pi)^2}, \qquad R= \frac{|\hat\chi|^{2/3}\zeta(3)^{2/3}}{2^{2/3}(2\pi)^2} - \frac{\chi_s}{3^{2/3} 12\kappa_{sss}^{1/3}},\qquad S=
\frac{2}{3^{2/3}a_s\kappa_{sss}^{1/3}}.
\end{equation}
We remind the reader that $\hat\chi$ is the Euler number of the CY 3-fold including an orientifold correction, $\kappa_{sss}$ and $\chi_s$ are the triple-self-intersection and Euler numbers of the small divisor and $a_s$ is the prefactor in front of the K\"ahler modulus in the exponent of the non-perturbative superpotential. The $\chi_s$ dependence is negligible when $\kappa_{sss}^{1/3}|\hat\chi|^{2/3}\gg 2.22\chi_s$ but relevant otherwise.

Using the above and assuming large $N=Q_3$, the leading behavior of the terms entering the bound \eqref{fb} is
\begin{align}
\hat\lambda^{(14)} &= \frac{3.95x^{1/3}}{a_s^{1/3}\kappa_{sss}^{1/9}|\hat\chi|^{2/9}Q_3^{1/3}}, \label{lambda14} \\
\hat\lambda^{(24)} &= \frac{6.62\cdot 10^2 \mathcal{W}_0\left(9.97\cdot 10^{-2}x^{4/7}a_s^{43/49}\kappa_{sss}^{43/147}|\hat\chi|^{41/147}Q_3^{3/7} \right)^2}{x a_s\kappa_{sss}^{1/3}|\hat\chi|^{2/3} Q_3}, \label{lambda24} \\
\hat\lambda^{(34)} &= \frac{1.72\cdot 10^2 \mathcal{W}_0\left(2.90 x^{2/5}a_s^{7/5}\kappa_{sss}^{7/15}|\hat\chi|^{-1/15}Q_3  \right)^2}{xa_s\kappa_{sss}^{1/3}|\hat\chi|^{2/3} Q_3}, \label{lambda34} \\
\hat\lambda^{(64)} &= \frac{0.169x^{1/3} a_s \kappa_{sss}^{4/9}|\hat\chi|^{2/9}}{Q_3^{1/3}} \label{lambda64}
\end{align}
with $x=1-2.22 \chi_s/\kappa_{sss}^{1/3}|\hat\chi|^{2/3}$.

Our bound thus constrains the topological invariants of the CY 3-fold and the orientifold/brane data of the compactification.
In particular, we observe that all four $\hat\lambda^{(i4)}$ terms become small as $Q_3$ is increased (note that the $1/Q_3$ factors in the second and third term win over the $Q_3$ dependence of the Lambert W-function). The LVS thus requires models with large tadpoles.
Large tadpoles were already proposed in \cite{Gao:2022fdi} to avoid the control issues of the LVS discovered in \cite{Junghans:2022exo}. However, as we will see momentarily, the bound \eqref{fb} puts much stronger constraints on the tadpole than the PTC of \cite{Gao:2022fdi} (see also Section \ref{sec:ptc0}). Another important observation is that \eqref{lambda14}--\eqref{lambda64} scale non-trivially with $a_s$, $\kappa_{sss}$, $\hat\chi$ and $\chi_s$. Whether or not an LVS model is under control does therefore not only depend on the tadpole but also on these other parameters. Note that $a_s$, $\kappa_{sss}$, $\hat\chi$ and $\chi_s$ appear in the numerators of some of the $\hat\lambda^{(i4)}$ terms and in the denominators of others. This implies that, at fixed $Q_3$, we cannot improve control indefinitely by dialing these parameters.

\begin{figure}[p]
\centering
\includegraphics[trim = 0mm 80mm 0mm 40mm, clip, width=0.8\linewidth]{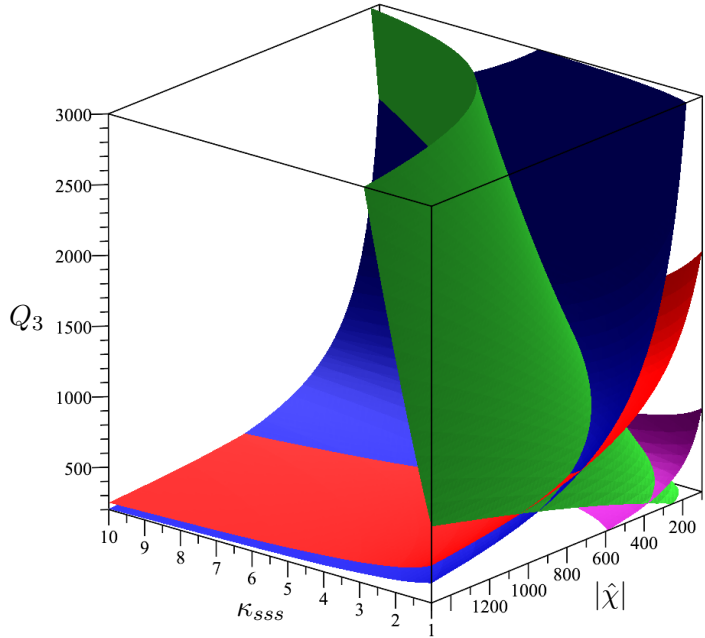}
\put(-324,135){$Q_3$}
\put(-238,25){$\kappa_{sss}$}
\put(-103,30){$|\hat\chi|$}

\caption{Constraints on $Q_3$, $\hat\chi$ and $\kappa_{sss}$ for $a_s=\frac{\pi}{3}$, $\chi_s=3$. The four surfaces demarcate the boundaries above which $\hat\lambda^{(14)}$ (red), $\hat\lambda^{(24)}$ (purple), $\hat\lambda^{(34)}$ (blue) and $\hat\lambda^{(64)}$ (green) are smaller than $0.1$. According to \eqref{fb}, dS vacua with $\lambda\le 0.1$ thus have to lie in the region above these four surfaces. This yields a minimally required $Q_3\sim 570$ if $\kappa_{sss}=1$ and $|\hat\chi|\sim 1040$. For different values of $\kappa_{sss}$ or $\hat\chi$, the required $Q_3$ quickly grows above $3000$.}
\label{NSR-plot}
\end{figure}

So far, our discussion was completely general. Let us now be more concrete. We first consider the case $a_s=\frac{\pi}{3}$, i.e., we consider models where the non-perturbative effect arises from gaugino condensation on a D7-brane stack with SO$(8)$ gauge group. We furthermore set $\chi_s=3$.
The constraints arising for this choice of $a_s$, $\chi_s$
are plotted in Fig.~\ref{NSR-plot}. As shown there, dS vacua with $\lambda\le 0.1$ would require at least a tadpole $Q_3\gtrsim 570$. This minimal value of the tadpole is consistent with our bound if $\kappa_{sss}=1$ and $|\hat\chi|\sim 1040$. For different $\kappa_{sss}$ and/or $\hat\chi$, the required tadpoles quickly rise to several thousand. We thus see that the result depends crucially on $\kappa_{sss}$ and $\hat\chi$.
Repeating the analysis for different values of $\chi_s$ leads to qualitatively similar conclusions. It is furthermore obvious that increasing the degree of control (e.g., demanding $\lambda\le 0.01$ instead of $\lambda\le 0.1$) would lead to even larger tadpoles than in Fig.~\ref{NSR-plot}.

\begin{figure}[p]
\centering
\includegraphics[trim = 0mm 80mm 0mm 40mm, clip, width=0.8\linewidth]{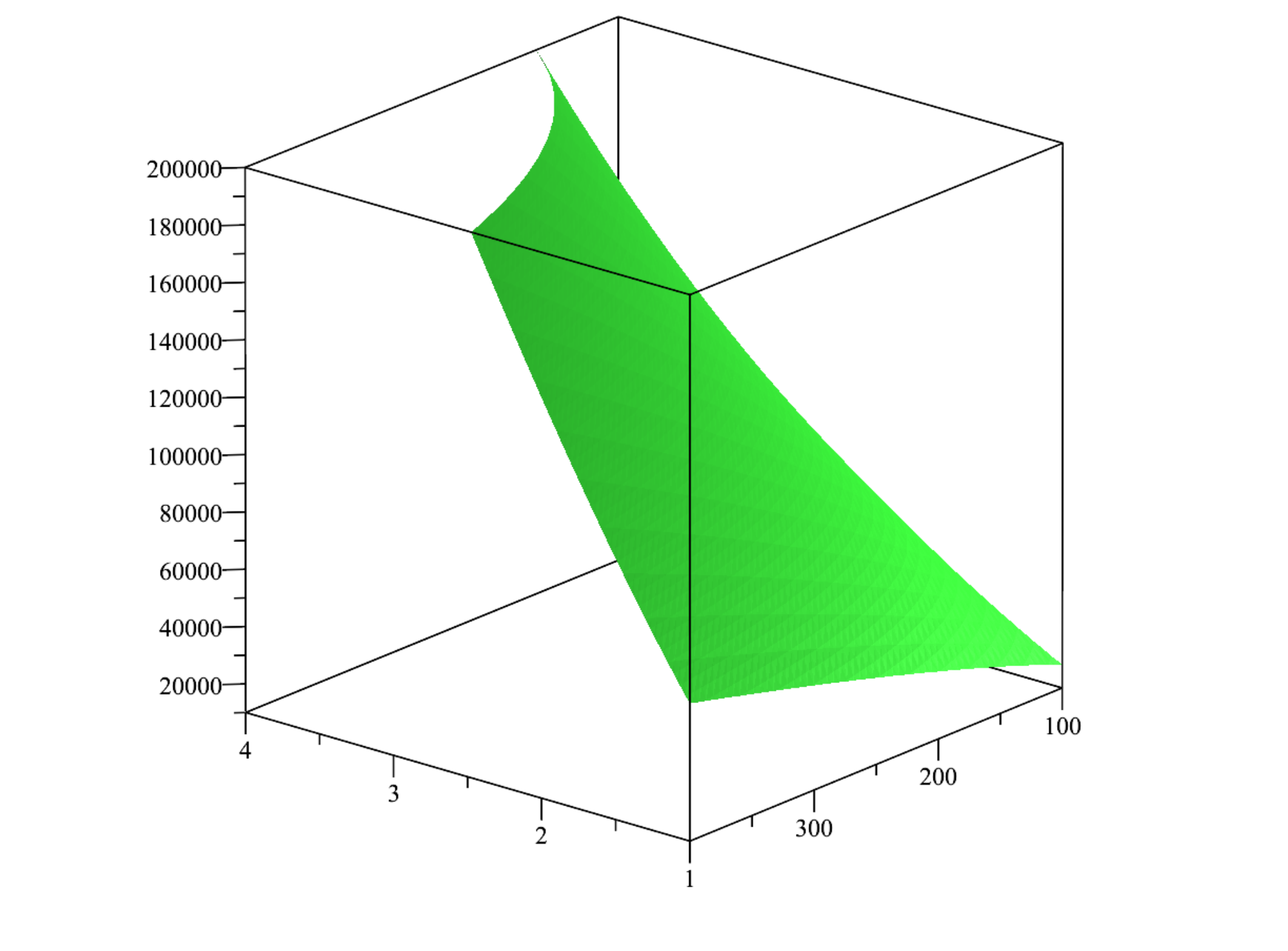}
\put(-330,135){$Q_3$}
\put(-238,25){$\kappa_{sss}$}
\put(-103,30){$|\hat\chi|$}

\includegraphics[trim = 0mm 80mm 0mm 40mm, clip, width=0.8\linewidth]{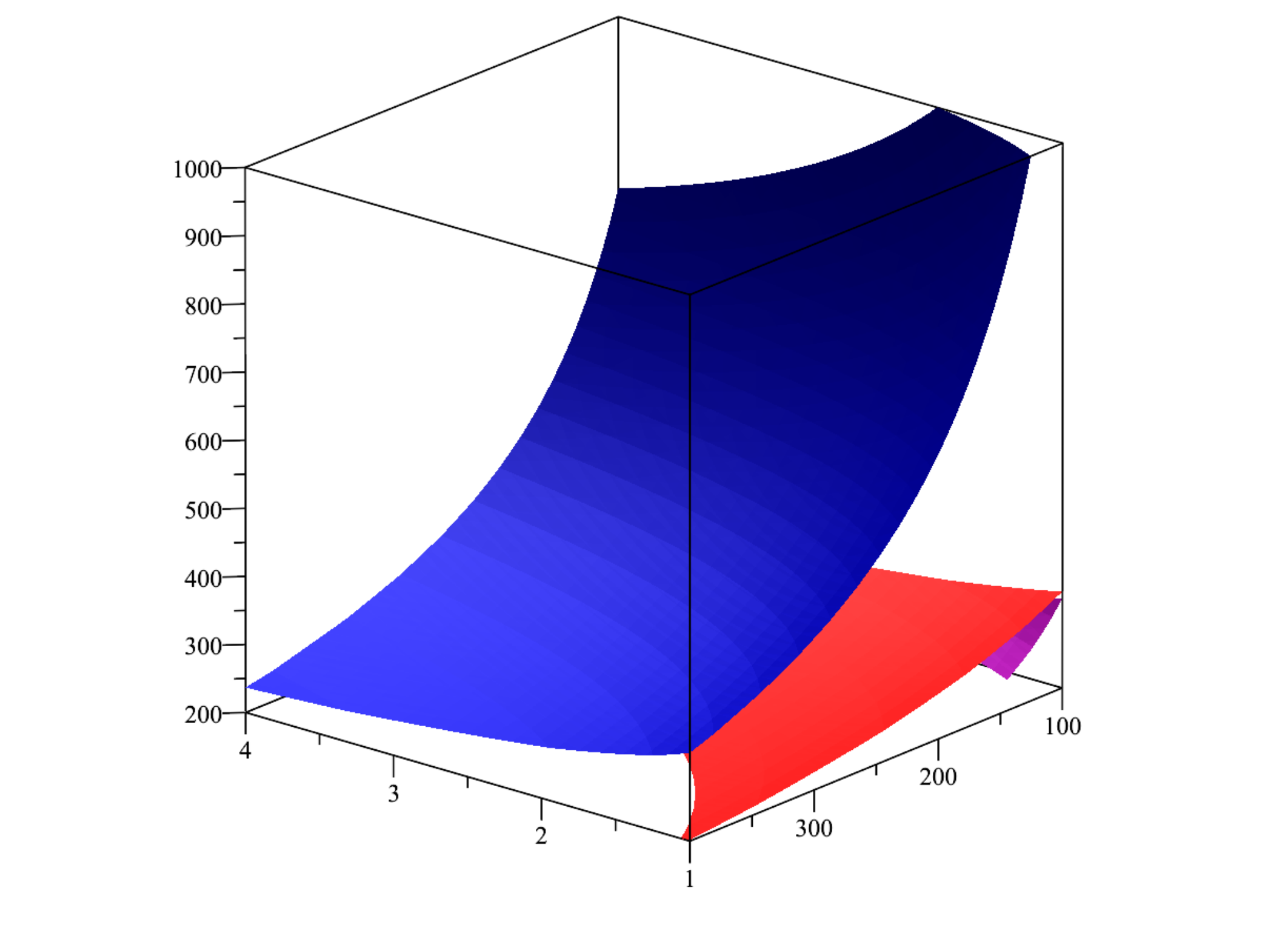}
\put(-322,135){$Q_3$}
\put(-238,25){$\kappa_{sss}$}
\put(-103,30){$|\hat\chi|$}
\caption{Constraints on $Q_3$, $\hat\chi$ and $\kappa_{sss}$ for $a_s=2 \pi$, $\chi_s=3$. The four surfaces demarcate the boundaries above which $\hat\lambda^{(14)}$ (red), $\hat\lambda^{(24)}$ (purple), $\hat\lambda^{(34)}$ (blue) and $\hat\lambda^{(64)}$ (green) are smaller than $0.1$. According to \eqref{fb}, dS vacua with $\lambda\le 0.1$ thus have to lie in the region above these four surfaces. Models with $a_s=2\pi$ therefore require huge tadpoles.}
\label{NSR-plot2}
\end{figure}

It is also interesting to consider the case $a_s=2\pi$, i.e., models where the non-perturbative effect is due to Euclidean D3 branes. In that case, our bound gets much stronger such that $\lambda\le 0.1$ requires $Q_3\gtrsim 20000$, see Fig.~\ref{NSR-plot2}. The reason is that, according to \eqref{lambda64}, $\hat\lambda^{(64)}$ is then $6$ times larger such that we require a $6^3=216$ times larger $Q_3$ to suppress it.

As a possible way out, one might wonder whether there are models in which some of the corrections we studied vanish, which would relax our constraints.
It is unclear to us why such a vanishing should happen, especially when supersymmetry is broken. However, one might perhaps imagine a situation where some of the numerical coefficients of the corrections, which are assumed to be $\mathcal{O}(1)$ in this paper, happen to be very small in a certain model and thus ``effectively'' vanish, either accidentally or by engineering it that way (for example, by stabilizing the complex-structure moduli in a favorable manner). Ensuring that this is the case would of course first require to compute the coefficients on smooth CYs, which is not possible with current technology. Also note that small coefficients are not expected to be generic but would correspond to a fine-tuning. One of the advantages of the LVS over the KKLT scenario (which requires a fine-tuned $W_0$) would thus be lost. Let us nevertheless entertain the possibility of such an ``LVS with mild corrections''.

For example, consider a (hypothetical) model in which no loop corrections are related to the small divisor (i.e., $\mathcal{C}_s^\text{KK}=0$ in the notation of \cite{Junghans:2022exo}). In such a model, the $\hat\lambda^{(64)}$ term in our bound \eqref{fb} would be absent. One then observes that the remaining three terms \eqref{lambda14}--\eqref{lambda34} contributing to the bound all become small for large $a_s$, $\kappa_{sss}$ and $|\hat\chi|$. For appropriate choices of these parameters, tadpoles of the order $Q_3\sim 200$ would thus be compatible with our bound, see Figs.~\ref{NSR-plot}, \ref{NSR-plot2}.\footnote{Note that, although suggested by \eqref{lambda14}--\eqref{lambda34}, one cannot decrease the required $Q_3$ indefinitely by considering larger and larger $\kappa_{sss}$, $|\hat\chi|$ beyond the ranges displayed in our plots. Indeed, one eventually reaches the $r\approx 0$ regime where the expansion used to compute the $\hat\lambda^{(i4)}$ terms in Section \ref{sec:bound} is not reliable anymore. In this regime, the $\hat\lambda^{(i4)}$ terms instead have to be computed using $g_s^{(i4)}\approx 40 c^\prime c^{\prime\prime}R/9 SN= 0.172xa_s\kappa_{sss}^{1/3}|\hat\chi|^{2/3}/N$. Doing so, one finds that the constraints on $Q_3$ are not relaxed at $r\approx 0$.} However, whether this is actually a viable route for model building is unclear to us.
Indeed, we traded the requirement of finding models with very large tadpoles for the (presumably even more difficult) requirement of computing loop coefficients in smooth CYs\footnote{Some contributions to $\mathcal{C}_s^\text{KK}$ can in principle be computed classically exploiting open/closed-string duality \cite{Junghans:2014zla}.} and finding models where they are very small or absent.

Let us finally discuss some recent advances in the literature on constructing models with large tadpoles.
See \cite{Carta:2020ohw, Altman:2021pyc, Crino:2022zjk} for explicit constructions of a large number of CY orientifolds including their tadpole data.
An idea emphasized, e.g., in \cite{Carta:2020ohw, Gao:2022fdi} and studied systematically in \cite{Crino:2022zjk} is that models with non-local D7 tadpole cancellation (i.e., models
where the D7 branes do not lie on top of the O7 planes but recombine into Whitney branes \cite{Aluffi:2007sx, Braun:2008ua, Collinucci:2008pf}) allow significantly larger D3 tadpoles than models where the D7 tadpole is cancelled locally (see also, e.g., \cite{Cicoli:2011qg, Louis:2012nb, Crino:2020qwk} for constructions exploiting this fact).
The largest tadpole found in \cite{Crino:2022zjk} is $Q_3=3332$ in a non-local model with $h^{1,1}_+=11$.\footnote{The tadpoles in \cite{Crino:2022zjk} are stated in the double cover of the orientifold so that we need to multiply them by a factor $1/2$.} If one restricts to $h^{1,1}_+=2$ (as in the minimal version of the LVS considered in this paper), the maximal tadpole found in \cite{Crino:2022zjk} is $Q_3=1839$. Models with even larger tadpoles were constructed in F-theory. In particular, the CY 4-fold with the largest known tadpole has $Q_3\approx 7.5\cdot 10^4$ \cite{Candelas:1997eh, Lynker:1998pb, Taylor:2015xtz}. If weakly coupled Swiss-cheese models with tadpoles of this order exist, one might hope that the control problems of the LVS described in this paper would be relaxed significantly. However, LVS models with non-local tadpole cancellation would face several obstacles, as we now explain:

\begin{itemize}
\item The examples with the largest tadpoles have a large number of K\"ahler moduli, i.e., one would have to go beyond the minimal version of the LVS with $h^{1,1}_+=2$. As already noted in \cite{Gao:2022fdi}, this would probably make it very challenging to construct \emph{explicit} dS vacua.
Already in the case of stabilizing just three moduli explicitly (i.e., two K\"ahler moduli and one conifold modulus), computing the solution, diagonalizing the mass matrix and studying the various corrections is a very difficult task \cite{Crino:2020qwk, Junghans:2022exo}.

\item Even a huge tadpole, say of the order $10^4$, would by itself not guarantee an absence of large corrections since our bound \eqref{fb} also depends strongly on other topological data, implying further non-trivial constraints.
For example, we observe from \eqref{lambda64} that the $\hat\lambda^{(64)}$ term is only suppressed at large $Q_3$ if $a_s^3\kappa_{sss}^{4/3}|\hat\chi|^{2/3}\lesssim Q_3$. Candidate models thus require a very large tadpole and at the same time a not-too-large $\kappa_{sss}$ and $|\hat\chi|$.
Another observation is that \eqref{lambda24}, \eqref{lambda34} are only suppressed when $\kappa_{sss}^{1/3}|\hat\chi|^{2/3}\gg 2.22\chi_s$. This is due to the fact that $\ln\mathcal{V} \sim a_s\tau_s-a_s\chi_s/24g_s$ in the LVS. The volume is therefore only exponentially large when $\chi_s$ is small enough.
On the other hand, it was emphasized in \cite{Gao:2022fdi, Crino:2022zjk} that large $Q_3$ requires the branes to wrap divisors with large Euler numbers (or an appropriately generalized notion thereof \cite{Aluffi:2007sx, Braun:2008ua, Collinucci:2008pf}). In a viable LVS model, this contribution has to almost exclusively come from the large divisor.

\item Models with non-local D7/O7 tadpole cancellation have the problem that the D7 branes and O7 planes backreact on the dilaton. This backreaction is expected to generate a correction at the order $g_s$ relative to the Euler-number term $\sim \hat\xi/g_s^{3/2}$ in the K\"ahler potential (see, e.g., the discussions in \cite{Minasian:2015bxa} around (3.9) and in \cite{Junghans:2022exo} in Section 3.1.3 and below (3.15)). An analogous correction arises in orbifold models, as can be shown by direct computation in string theory \cite{Antoniadis:2019rkh}.\footnote{Contrary to the smooth CY case, the Euler-number term in orbifold models arises at the one-loop order. The correction computed in \cite{Antoniadis:2019rkh} is due to a genus-$3/2$ diagram and is thus suppressed by a power of $g_s$ relative to the Euler-number term as in the smooth case.} As shown in \cite{Junghans:2022exo}, these backreaction corrections (labelled by $\mathcal{C}_1^\xi$, $\mathcal{C}_2^\xi$ there) lead to a loss of control in the LVS, as they blow up at small coupling in the exponent of the volume vev. It is unclear how to avoid this in non-local models.

\end{itemize}
We thus see that it is far from obvious whether models with non-local tadpole cancellation can solve the control problems of the LVS. On the other hand, models with local tadpole cancellation satisfy $Q_3\le 252$ (see, e.g., \cite{Gao:2022fdi, Crino:2022zjk}), which is significantly below the bound $Q_3\gtrsim 570$ found at the beginning of this section. For these reasons, it appears very challenging to construct LVS dS vacua in a controlled way.

\subsection{How tight is the bound?}
\label{sec:disc2}

We have seen that the bound \eqref{fb} puts strong constraints on the topology and orientifold/brane data of LVS models.
As we now explain, the true control problems are actually even worse than indicated by this bound. Indeed, one can show that the bound cannot be saturated, i.e., the value of $\lambda$ in the best-controlled dS solutions is strictly larger than the right-hand side of \eqref{fb}. This in turn implies that achieving a certain level of control (say $\lambda\le 0.1$ as in the previous section) requires the tadpoles to be even larger than predicted by \eqref{fb}, i.e., controlled dS vacua lie well above the surfaces in Figs.~\ref{NSR-plot}, \ref{NSR-plot2} rather than directly on top of them.
The reason is that our derivation in Section \ref{sec:bound} ignores the following three effects:
\begin{itemize}
\item 
We used that the tadpole $N=KM$ induced by the conifold fluxes is bounded by $N\le Q_3$. However, in a realistic model, one should not use the full available tadpole for the conifold fluxes but rather leave room for further fluxes to be able to stabilize the complex-structure/D7-brane moduli. It is difficult to estimate in general how much these fluxes contribute to the tadpole. However, recent results suggest that this contribution grows faster than $Q_3$ as the number of complex-structure/D7-brane moduli becomes large \cite{Bena:2020xrh, Grana:2022dfw} (unless one considers singular compactifications \cite{Gao:2022fdi}).
Indeed, for $h_+^{1,1}=2$, large $Q_3$ requires a large number of complex-structure/D7-brane moduli (see, e.g., Section 5 in \cite{Gao:2020xqh}). Hence, already for moderately large $Q_3$, we expect that only a small fraction of $Q_3$ is available to be used for the conifold fluxes, i.e., $N \ll Q_3$.
Even just reserving half of the tadpole for moduli stabilization, i.e., $N\le \frac{1}{2}Q_3$, our bounds on $Q_3$ in the previous section would become twice as strong.

\item The two inequalities \eqref{bb2}, \eqref{bb3} leading to our bound \eqref{fb} cannot always be saturated at the same time. To see this, recall that \eqref{bb2} is obtained
by minimizing each of the $\bar\lambda_i$ parameters separately with respect to $\alpha$.
According to \eqref{hatalpha}, these minima lie at a different value of $\alpha$ for each $\bar\lambda_i$ parameter.
It follows that, regardless of the value $\alpha$ takes in a given solution, some of the $\bar\lambda_i$ parameters sit away from their minimum, i.e., $\bar\lambda_i>\hat\lambda_i$ for some $i$.
On the other hand, \eqref{bb3} is saturated if and only if $\hat\lambda=\hat\lambda_4=\hat\lambda_i$ where $i$ equals $1$, $2$, $3$ or $6$. Let us for example assume $i=1$ so that $\hat\lambda=\hat\lambda_4=\hat\lambda_1$.\footnote{An analogous argument would apply for $i=2,3$ but not in the case $i=6$ since $\bar\lambda_6$ does not depend on $\alpha$.} According to the above discussion, we also have either $\bar\lambda_4>\hat\lambda_4$ or $\bar\lambda_1>\hat\lambda_1$. It follows that $\bar\lambda>\hat\lambda$ and so \eqref{bb2} is not saturated, which confirms our initial claim.

\item We assumed in Section \ref{sec:bound} that $W_0$, $\alpha$, $g_s$ and $N$ can be chosen in such a way that the corrections are minimized. Let us call this choice the ``optimal choice''.
What this neglects is that $W_0$, $\alpha$, $g_s$ and $N$ have to satisfy the constraint \eqref{alpha}, which after substituting the moduli vevs and taking the log yields
\begin{align}
\frac{8\pi k}{5}&=\frac{a_s\hat\xi^{2/3}}{(2\kappa_s)^{2/3}} -\frac{a_s\chi_s}{24}
+ g_s\left(\frac{1}{3}+\frac{4\alpha}{15} -\frac{3}{5}+\frac{3}{5}\sqrt{1-\frac{64\pi c^\prime c^{\prime\prime} k}{9g_sN}}\right) \nl
+g_s\ln\left(\frac{3\cdot 40^{3/5}q_0^{3/5} (2\kappa_s)^{4/15}\hat\xi^{2/15}}{8\cdot 27^{3/5}\alpha^{3/5} a_s^{2/5} g_s^{4/5} |W_0|^{1/5}} \right) +\mathcal{O}(g_s^2). \label{xx1}
\end{align}
Note that this is the same constraint as \eqref{xx0}, \eqref{xx} but including the next-to-leading $g_s$ corrections.

We thus see that $k\equiv K/M$ is fixed in terms of $W_0$, $\alpha$, $g_s$, $N$ and the various topological parameters. This is problematic since $k$ is not an arbitrary real number but heavily restricted by the requirements of flux quantization and tadpole cancellation. In particular, the allowed choices for $K$ and $M$ are given by the factorizations of $N$ into two integers. The average number of such factorizations is rather small even for large $N$. For example, in the regime $\mathcal{O}(10^2)\lesssim N \lesssim \mathcal{O}(10^3)$, the average number of integer factorizations is only $\mathcal{O}(10)$. For a given $N$, we therefore typically have very few possibilities to choose $K$ and $M$ and can thus only adjust the left-hand side of \eqref{xx1} very crudely. This means that \eqref{xx1} is generically violated if we take the optimal choice for $W_0$, $g_s$, $\alpha$, $N$ that went into the derivation of our bound in Section \ref{sec:bound}.

Let us make a (very rough) estimate of how large this violation is on average, i.e., how large the ``steps'' $\Delta k$ are in which we can vary the discrete parameter $k$.
To this end, we assume that $N$ lies in an interval $[Q_3-q,Q_3]$. Note that $q$ should not be too large since we want $N$ to be as large as possible for optimal control. $K/M$ can take values in $[Q_3^{-1},Q_3]$, and
we have $\mathcal{O}(10)(q+1)$ different values of $K/M$ in this interval. The average step size is thus
\begin{equation}
\Delta k = \frac{Q_3-Q_3^{-1}}{\mathcal{O}(10)(q+1)}\simeq \frac{Q_3}{\mathcal{O}(10)(q+1)}. \label{delta}
\end{equation}
For small $(q+1)/Q_3$, this yields $\Delta k \gtrsim\mathcal{O}(1)$. Hence, flux quantization and tadpole cancellation imply that $k$ is typically off by $\Delta k\gtrsim\mathcal{O}(1)$ from the value that would be required to solve \eqref{xx1} with the optimal choice for $W_0$, $g_s$, $\alpha$, $N$.
We are thus forced to choose values of $W_0$, $\alpha$, $g_s$, $N$ that significantly differ from the optimal choice assumed for our bound. It is furthermore evident from \eqref{xx1} that changing $g_s$ and $\alpha$ does by itself not help much, as this only gives a small correction unless $\alpha$ leaves the dS range \eqref{range} or $g_s$ is taken large. We thus need to move $W_0$ and/or $N$ away from their optimal choice.

Let us discuss the first possibility first, i.e., we compensate for the shift $\Delta k$ in \eqref{xx1} by a corresponding rescaling of $W_0$ away from its optimal value, $W_0 \to W_0 \exp(-\frac{8\pi \Delta k}{g_s})$.
On the other hand, recall that we assumed in Section \ref{sec:bound} that we can always adjust $W_0$ in such a way that $\bar\lambda_5$ is not larger than the other $\bar\lambda_i$ parameters and is thus not relevant for our bound.
One can furthermore check that being able to choose $W_0$ freely also allows to adjust $\bar\lambda_5$ such as to saturate the inequality \eqref{bb1} used for the derivation of our bound \eqref{fb}.
However, if $W_0$ differs from this optimal choice by an exponentially large factor $\exp(-\frac{8\pi \Delta k}{g_s})$, these arguments no longer apply
and, consequently, the smallest possible $\lambda$ in a given model can be much larger than our bound.

The second possibility is to look for a number $N$ with an integer factorization such that $\Delta k\ll 1$, i.e., such that $k$ is very close to the value needed
to solve \eqref{xx1} with optimal $W_0$, $\alpha$, $g_s$.
We can increase our chances of finding such an integer factorization by scanning through many different values of $N$. According to our estimate \eqref{delta}, this requires a large $q=\mathcal{O}(Q_3)$, i.e., we generically have to include values of $N$ in the search that lie significantly below $Q_3$ until we find a favorable choice. Hence, even assuming that moduli stabilization does not eat up much of the tadpole (see bullet point 1), the required choice for $N$ may turn out to be much smaller than the naively best choice $N=Q_3$
because only very few numbers have integer factorizations such that \eqref{xx1} is solved with nearly optimal $W_0$, $\alpha$, $g_s$.
For example, in the analysis of \cite{Junghans:2022exo} (which was specific to the model \eqref{crino}), it turned out that the best-controlled dS solutions have $N=115$ even though naively the solutions with the maximal $N=Q_3=149$ should be better controlled. This is precisely due to the fact that $N=115$ admits a factorization $K=5$, $M=23$ which allows to satisfy \eqref{xx1} with values of $W_0$, $\alpha$ and $g_s$ that are close to optimal.

\end{itemize}

To summarize, we discussed several effects that further strengthen the bound on $\lambda$ compared to our result in Section \ref{sec:bound}.
The tadpoles required in LVS models are therefore even larger than the already quite large tadpoles implied by our bound.
It would be very interesting to quantify precisely how large the neglected effects are. Importantly, they all \emph{increase} the bound.\footnote{Our expressions in Section \ref{sec:bound} also receive subleading $g_s$ corrections. These can a priori increase or decrease the bound compared to the leading result but, contrary to the other effects, they become small at large $N$.}
However, it is technically involved to systematically take this into account in a model-independent way. For example, this seems to require going through each possible integer factorization for all $N$ by hand (as we did in \cite{Junghans:2022exo} for the model \eqref{crino}). We have refrained from performing such a tedious analysis in this work, as the bound we derived here is already very strong without taking into account the above effects.

\section{Conclusions}
\label{sec:concl}

The main result of this paper is the inequality \eqref{fb}, which together with \eqref{hl2}, \eqref{hl3}, \eqref{hl1} and \eqref{hl6} provides a bound on how well perturbative corrections can in principle be controlled in LVS dS vacua. Our bound is model-independent and in particular holds for any choice of CY orientifold (assuming the minimal scenario with $h_+^{1,1}=2$ and an uplift generated by an anti-D3 brane). A consequence of our bound is that the topology and the orientifold/brane data of candidate LVS models are strongly constrained, as illustrated in Figs.~\ref{NSR-plot}, \ref{NSR-plot2}. In particular, we found that such models would require a very large D3 tadpole. The minimally required tadpole for $\lambda\le 0.1$ ranges from $\mathcal{O}(500)$ to $\mathcal{O}(10^6)$ or more and depends crucially on the Euler number of the CY 3-fold, the triple-self-intersection and Euler numbers of the small divisor and the coefficient $a_s$ appearing in the non-perturbative superpotential. We also discussed several effects that we neglected in our analysis and that further increase the required tadpole compared to our already very strong bound.

Whether it is possible to construct models that satisfy these strong constraints remains to be seen. We gave arguments suggesting that this may not be possible, in spite of recent interesting advances \cite{Crino:2022zjk} in constructing models with large D3 tadpoles. In particular, the latter have D7 tadpoles which are cancelled non-locally. However, as observed in \cite{Junghans:2022exo}, such non-local models are expected to suffer, among other possible problems, from large backreaction corrections that scale like $1/g_s$ in the exponent of the volume and thus lead to a loss of control at small coupling. On the other hand, models with local D7-tadpole cancellation have relatively small D3 tadpoles and can thus probably not satisfy our constraints.

Aside from the corrections studied in this paper, a further type of potentially dangerous corrections are generated by 1-loop redefinitions of the K\"ahler moduli \cite{Conlon:2009xf, Conlon:2009kt, Conlon:2010ji}. As emphasized in \cite{Junghans:2022exo}, these corrections also blow up at small $g_s$, similar to the backreaction corrections mentioned above. Finding models in which these 1-loop redefinitions are absent would present an additional obstacle beyond satisfying the strong topological constraints derived in this paper.
Unless one can establish the existence of models satisfying all these requirements or develop techniques to systematically compute all unsuppressed corrections, LVS dS vacua cannot be self-consistently constructed and would thus be in the swampland.

\section*{Acknowledgments}

I would like to thank the authors of \cite{Gao:2022fdi} for a stimulating correspondence.

\appendix

\section{Comparison of PTC versions}
\label{sec:ptc}

The coefficient $\gamma^\prime$ in our version of the PTC differs from the corresponding coefficient appearing in (3.10)--(3.12) of \cite{Gao:2022fdi}. This is mainly due to different definitions of the anti-brane potential and the control parameters in the two papers, as we will now explain in more detail.
We first discuss the definition of the anti-brane potential:
\begin{itemize}
\item In the approach of \cite{Crino:2020qwk} adopted here and in \cite{Junghans:2022exo}, the anti-brane is incorporated into the $\mathcal{N}=1$ supergravity EFT using a nilpotent superfield and keeping the conifold modulus dynamical \cite{Douglas:2007tu, Douglas:2008jx, Bena:2018fqc, Blumenhagen:2019qcg, Dudas:2019pls}, whereas \cite{Gao:2022fdi} uses the traditional approach of adding the uplift term to the potential by hand. The latter approach does not include the backreaction of the anti-brane uplift on the $\zeta$ vev.
This corresponds to expanding $\zeta$ and $q_0$ in $1/g_sM^2$ and only keeping the leading terms. Indeed, one can verify that the parametric scalings of our expressions then agree with those in \cite{Gao:2022fdi}.
Note, however, that, for smaller $g_sM^2$ close to the bound $g_sM^2\approx 46.1$, backreaction effects in $\zeta$ cannot be neglected. On the other hand, the approach using a dynamical $\zeta$ has some caveats as well, see, e.g., Section 2 in \cite{Junghans:2022exo} or \cite{Gao:2020xqh} for a discussion.
The bound computed in Section \ref{sec:bound} arises in the regime $\frac{40 c^\prime c^{\prime\prime}R}{9g_sSN}\ll 1$ and thus, by \eqref{nk} and \eqref{xx}, in the regime of large $g_sM^2$ where the backreaction on $\zeta$ is negligible.

\item 

It was pointed out in \cite{Junghans:2022exo} that the conifold/anti-brane potential (i.e., the second line in \eqref{lvs-potential}) has an ambiguity related to the gluing of the Klebanov-Strassler (KS) throat \cite{Klebanov:2000hb} into the compact CY manifold.
To see this, note that the internal (Einstein-frame) metric can be written as $g_{mn}=\e^{-2A}\tilde g_{mn}$, where $\tilde g_{mn}$ is Ricci flat in the weak-coupling limit and $\e^{-4A}=\e^{-4A_0}+c$ is the warp factor, which we split into a varying part $\e^{-4A_0}$ and a constant $c$ specified below \cite{Giddings:2001yu, Giddings:2005ff}. We further choose to normalize $\e^{-4A_0}$ and $\tilde g_{mn}$ such that $\e^{-4A_0}\to \e^{-4A_\text{KS}}$, $\e^{-2A_0}\tilde g_{mn}\to g_{mn}^\text{KS}$ as we enter the throat region, where we denote by ``KS'' the (Einstein-frame) KS solution \cite{Klebanov:2000hb, Herzog:2001xk}. Moving away from the throat region, $\e^{-4A_0}$ and $\tilde g_{mn}$ start to deviate from the KS solution, and corrections that depend on the bulk geometry become larger. This implies in particular that the volume of the unwarped metric, $\mathcal{\tilde V} \equiv \int_X \d^6 y \sqrt{\tilde g}$, is a model-dependent constant which is not fixed by the local throat geometry, as it involves an integration over the whole bulk.
Furthermore, the constant $c$ is an integration constant at the level of the 10d supergravity equations \cite{Giddings:2005ff} and related to the volume modulus as follows. We define $\mathcal{V}$ as the volume measured with $g$, i.e., $\mathcal{V} \equiv \int_X \d^6 y \sqrt{g} \approx c^{3/2}\mathcal{\tilde V}$, where we substituted the metric and the warp factor and assumed the large-$c$/large-volume regime in the last step. We thus find $c = \mathcal{V}^{2/3}/\mathcal{\tilde V}^{2/3}$ and therefore $\e^{-4A}=\e^{-4A_0}+\mathcal{V}^{2/3}/\mathcal{\tilde V}^{2/3}$.

Let us now switch to the usual convention where $\e^{-4A}\to 1$ in the large-volume limit, which can be achieved by absorbing an appropriate rescaling of $\e^{-4A}$ into $\tilde g_{mn}$.
The warp factor in the strongly-warped throat region is then
\begin{equation}
\e^{-4A}= 1 + \frac{\mathcal{\tilde V}^{2/3} \e^{-4A_0}}{\mathcal{V}^{2/3}} \approx \frac{\mathcal{\tilde V}^{2/3} \e^{-4A_\text{KS}}}{\mathcal{V}^{2/3}}. \label{aks}
\end{equation}
We thus see that the warp factor in the throat depends on the model-dependent constant $\mathcal{\tilde V}$. This is relevant for the uplift since the conifold/anti-brane part of the potential is proportional to $\e^{4A}$ \cite{Douglas:2007tu, Douglas:2008jx, Bena:2018fqc, Blumenhagen:2019qcg, Dudas:2019pls} and thus sensitive to $\mathcal{\tilde V}$.\footnote{The same potential could also be derived choosing a different normalization of the warp factor. Indeed, the warp-factor normalization cancels in the potential with a corresponding rescaling factor from transforming to 4d Einstein frame.} As noted in \cite{Junghans:2022exo}, this ambiguity can be taken into account by introducing a constant $\gamma_0\sim g_s\mathcal{\tilde V}^{2/3}$ in the K\"ahler potential. This yields, e.g., an anti-brane term
\begin{equation}
V_{\overline{\text{D}3}} = \frac{c^{\prime\prime}\zeta^{4/3}}{2\pi \gamma_0 g_sM^2 \mathcal{V}^{4/3}} \label{ab}
\end{equation}
and an analogous $\gamma_0$ scaling of the other terms in the second line of \eqref{lvs-potential}.
Note that $\gamma_0$ was not discussed in previous works such as \cite{Crino:2020qwk} and implicitly set to 1 there.
On the other hand, an attempt to estimate $\gamma_0$ was made in \cite{Gao:2022fdi}. Comparing the above potential to (A.10) there, we find $\gamma_0\approx 0.0235$ and thus an about 40 times larger uplift term compared to the assumption here and in previous works.

In this work, we stick to the choice $\gamma_0=1$ following \cite{Junghans:2022exo}, as it is not clear to us how to determine $\gamma_0$ reliably without performing the gluing in an explicit geometry and computing the metric of the resulting compact CY. Further note that, as explained in \cite{Junghans:2022exo}, $\gamma_0$ appears in the denominator of some of the $\lambda_i$ parameters and in the numerator of others such that its value cannot significantly affect the conditions for control. In particular, it follows from Section 2 in \cite{Junghans:2022exo} that the $\gamma_0$ dependence in any of our expressions can be reinstated by replacing $c^{\prime}\to c^{\prime}\gamma_0$, $c^{\prime\prime}\to c^{\prime\prime}/\gamma_0$. Applying this to the $\bar\lambda_i$ parameters in Section \ref{sec:bound}, we find
\begin{equation}
\bar\lambda_1 \sim \bar\lambda_3 \sim \bar\lambda_4\sim \bar\lambda_5\sim \bar\lambda_6 \sim \gamma_0^0, \qquad \bar\lambda_2 \sim \gamma_0^{4/3}.
\end{equation}
A very small $\gamma_0$ as in \cite{Gao:2022fdi} would therefore not affect the bound derived in Section \ref{sec:bound}, which is due to $\bar\lambda_1$, $\bar\lambda_3$, $\bar\lambda_4$ and $\bar\lambda_6$. However, sending $c^{\prime}\to c^{\prime}\gamma_0$ implies $q_0\to q_0/\gamma_0$ and thus modifies, at next-to-leading order in $g_s$, the constraint \eqref{xx1} (and \eqref{kc2} below) relating $k\equiv K/M$ to the other parameters.

\end{itemize}

Aside from the anti-brane potential, another difference between our formalism and the one in \cite{Gao:2022fdi} is the definition of the parameters that control the warping and higher $F$-term corrections. These parameters are denoted by $\lambda_3$, $\lambda_5$ here and by $1/c_N$, $1/c_{W_0}$ in \cite{Gao:2022fdi} and satisfy
\begin{equation}
\lambda_3\sim \frac{1}{c_N}\sim \frac{KM}{g_s\mathcal{V}^{2/3}}, \qquad \lambda_5\sim \frac{1}{c_{W_0}}\sim \frac{|W_0|^2}{\mathcal{V}^{2/3}}
\end{equation}
with different proportionality factors in the two papers, which
are explained as follows:
\begin{itemize}
\item In the approach of \cite{Junghans:2022exo}, which we follow here, the control parameters are defined based on the requirement that the corrections are small in two quantities called $\rho$ and $\mu_3$, which are proportional to the vacuum energy and one of the eigenvalues of the Hessian, respectively. On the other hand, \cite{Gao:2022fdi} defines the control parameters based on the requirement that the corrections in the potential are small compared to the AdS part of the leading potential. Note that both requirements are necessary conditions for control.

\item The definition of $1/c_N$ in \cite{Gao:2022fdi} differs from our $\lambda_3$ by a further factor $\hat\xi$.
As shown in \cite{Junghans:2022exo}, $\lambda_3 \sim \int_X \e^{-4A_0}c_3(X)/\mathcal{V}^{2/3}$, where $c_3$ is the Euler class and $\e^{-4A_0}$ denotes the varying part of the warp factor as above.
The extra factor $\hat\xi$ included in \cite{Gao:2022fdi} arises from the assumption that the integral $\int_X \e^{-4A_0}c_3(X)$ is proportional to the Euler number.  However, it is not obvious why this would be the case since $\e^{-4A_0}$ is a non-trivial function that varies over the internal space. Note that $\e^{-4A_0}$ does not contain any implicit volume dependence and in particular does not approach a constant in the large-volume limit.
Including nevertheless the extra $\hat\xi$ factor in the definition of $\lambda_3$, the bound derived in Section \ref{sec:bound} would become even stronger in the regime of large Euler numbers $\hat\xi > 1$.

\end{itemize}

Let us finally also discuss an ambiguity in how $\hat\xi$ is related to the conifold fluxes at weak coupling. In particular, taking the log of \eqref{alpha} and using the moduli vevs and the definition of $\lambda_5$ stated in Section \ref{sec:review}, one finds
\begin{align}
\frac{a_s\hat\xi^{2/3}}{(2\kappa_s)^{2/3}g_s} -\frac{a_s\chi_s}{24g_s} &=
\frac{8\pi N}{5g_sM^2} - \frac{1}{3}-\frac{4\alpha}{15} +\frac{3}{5}-\frac{3}{5}\sqrt{1-\frac{64\pi c^\prime c^{\prime\prime}}{9g_sM^2}} +\mathcal{O}(g_s) \nl
+\ln\left(\frac{8\cdot 27^{3/5}\alpha^{3/5} a_s^{2/5} g_s^{4/5} |W_0|^{1/5}}{3\cdot 40^{3/5}q_0^{3/5} (2\kappa_s)^{4/15}\hat\xi^{2/15}} \right)
 \nll
= \frac{16\pi N}{9g_sM^2} - \frac{1}{3}-\frac{4\alpha}{15} +\frac{2}{3}-\frac{2}{3}\sqrt{1-\frac{64\pi c^\prime c^{\prime\prime}}{9g_sM^2}} +\mathcal{O}(g_s) \nl
+ \ln \left( \frac{8\cdot 27^{5/6}a_s^{1/6}g_s^{5/6} \alpha^{2/3} \lambda_5^{1/6}}{3\cdot 40^{2/3} q_0^{2/3} \text{max}\left(\frac{32}{|\alpha-1|},\frac{88}{|\frac{9}{4}-\alpha|}\right)^{1/6}\hat\xi^{1/18}(2\kappa_s)^{1/9}} \right). \label{kc2}
\end{align}
Assuming small $g_s$, it follows that $\frac{a_s\hat\xi^{2/3}}{(2\kappa_s)^{2/3}} -\frac{a_s\chi_s}{24} \approx \frac{8\pi N}{5M^2}$ if the argument of the log in the second line is $\mathcal{O}(1)$, and that $\frac{a_s\hat\xi^{2/3}}{(2\kappa_s)^{2/3}} -\frac{a_s\chi_s}{24} \approx \frac{16\pi N}{9M^2}$ if the argument of the log in the last line is $\mathcal{O}(1)$.
We assume that the first relation holds in this paper, while \cite{Gao:2022fdi} assumes the second relation (and sets $\chi_s=0$). Note that it does not make much of a difference which of the two relations one imposes, as the ratio between the two terms on the right-hand sides is $\frac{10}{9}$.

\bibliographystyle{utphys}
\bibliography{groups}

\end{document}